\documentclass[a4paper,fleqn,usenatbib]{mnras}
\usepackage{newtxtext,newtxmath}

\usepackage[T1]{fontenc}
\usepackage[english]{babel}

\usepackage{caption} 
\usepackage{float}   
\usepackage{graphicx}

\newcommand\mean[1]{\overline{#1}}
\newcommand*{\swap}[2]{#2#1} 

\def\co{${}^{56}$Co}
\def\ni{${}^{56}$Ni}
\def\ergs{erg\,s$^{-1}$}

\def\cm3{cm$^{-3}$}
\def\kms{km~s$^{-1}$}
\def\lsun{L$_{\odot}$}
\def\rsun{R$_{\odot}$}
\def\msun{M$_{\odot}$}
\def\one{{\,\sc i}}
\def\two{{\,\sc ii}}

\newcommand{\iso}[2]{\ensuremath{^{#1}\rm{#2}}}

\def\v1d{{\sc v1d}}

\def\mesa{{\sc mesa}}
\def\cmfgen{{\sc cmfgen}}

\title[Progenitors of low-luminosity Type II-Plateau supernovae]
      {Progenitors of low-luminosity Type II-Plateau supernovae}
\author[S.~M.~Lisakov et al.]
{Sergey~M.~Lisakov,$^1$\thanks{E-mail:
  \href{mailto:lisakov57@gmail.com}{lisakov57@gmail.com} (SML)}
Luc Dessart,$^{2,1}$\thanks{E-mail:
  \href{mailto:Luc.Dessart@oca.eu}{Luc.Dessart@oca.eu} (LD)}
D. John Hillier,$^{3}$
Roni Waldman,$^4$
and Eli Livne$^4$
  \\
  \\
$^1$: Laboratoire Lagrange, UMR7293, Universit\'e Nice Sophia-Antipolis,
Observatoire de la C\^{o}te d'Azur, 06304 Nice, France. \\
$^2$: Unidad Mixta Internacional Franco-Chilena de Astronom\'ia (CNRS UMI
3386), Departamento de Astronom\'ia, Universidad de Chile,\\
Camino El Observatorio 1515, Las Condes, Santiago, Chile\\
$^3$: Department of Physics and Astronomy \& Pittsburgh Particle Physics,
Astrophysics, and Cosmology Center (PITT PACC),  University of Pittsburgh, \\
3941 O'Hara Street, Pittsburgh, PA 15260, USA. \\
$^4$: Racah Institute of Physics, The Hebrew University, Jerusalem 91904, Israel. \\
}

\begin{document}
\pagerange{\pageref{firstpage}--\pageref{lastpage}} \pubyear{2017}
\maketitle
\label{firstpage}

\begin{abstract}
The progenitors of low-luminosity Type II-Plateau supernovae (SNe II-P) are
believed to be red supergiant (RSG) stars, but there is much disparity in the literature
concerning their mass at core collapse and therefore on the main sequence.
Here, we model the SN radiation arising from the low-energy explosion of RSG
stars of 12, 25, and 27\,\msun\ on the main sequence and formed through single star evolution.
Despite the narrow range in ejecta kinetic energy (2.5$-$4.2$\times$10$^{50}$\,erg)
in our model set, the SN observables from our three models are significantly distinct,
reflecting the differences in progenitor structure (e.g., surface radius, H-rich envelope mass,
He-core mass). Our higher mass RSG stars give rise to Type II SNe
that tend to have bluer colors at early times, a shorter photospheric phase, and
a faster declining $V$-band light curve (LC) more typical of Type II-linear SNe,
in conflict with the LC plateau observed for low-luminosity SNe II.
The complete fallback of the CO core in the low-energy explosions of our high mass RSG
stars prevents the ejection of any \ni\ (nor any core O or Si), in contrast to low-luminosity
SNe II-P, which eject at least 0.001\,\msun\ of \ni.
In contrast to observations, type II SN models from higher mass RSGs tend to show 
an H$\alpha$ absorption that remains broad at late times (due to a larger velocity 
at the base of the H-rich envelope).
In agreement with the analyses of pre-explosion photometry, we conclude that low-luminosity
SNe II-P likely arise from low-mass rather than high-mass RSG stars.
\end{abstract}


\section{Introduction}\label{intro}

Type II supernovae (SNe) are thought to arise from stars with an initial mass between
$\sim$\,8 and $\sim$\,30\,\msun, that end their lives in the red-supergiant (RSG) stage
with the collapse of their degenerate core
\citep{Woosley1995, Heger2003, Limongi2003,Eldridge2004, Hirschi2004}.
Historically, Type II SNe have been divided into Type
II-Plateau (II-P) and Type II-Linear (II-L) based on their light curve (LC)
morphology \citep{Barbon1979}. More recent surveys that collected tens to hundreds
of SNe II find a continuous distribution of decline rates in $V$-band
LCs, which suggests that the division between II-P and II-L is somewhat
artificial \citep{Anderson2014a, Sanders2015}.
RSG progenitors that kept a sufficient amount of hydrogen
in the envelope yield a Type II SN with an optical-brightness plateau for 3--4 months
and prominent H\one\ lines in their spectra. The plateau duration is controlled
by the mass of the H-rich envelope and the surface radius,
as well as the explosion energy and the \ni\ mass
(e.g., \citealp{Grassberg1971, Falk1977, Litvinova1983, Swartz1991, Popov1993}).
Type II SNe that show a fast declining $V$-band LC likely arise from RSG progenitors
that have a lower H-rich envelope mass (\citealt{BB_2l_92}, \citealt{moriya_2l_16};
but see \citealt{nakar_snii_16}).

The class of Type II-P SNe is rather homogeneous in terms
of plateau duration and spectral features, but the plateau luminosity
may vary over a wide range.
Indeed, over the last $\sim$\,20 years a number of faint Type II SNe have been
observed. Spectroscopic and photometric observations for most of these
objects have been presented and analyzed in \citet{Pastorello2004}
and \citet{Spiro2014}. These works emphasize the distinctive features of
low-luminosity Type II SNe:

\begin{enumerate}
\item a low expansion rate (approximately in the range
from 1300 to 2500~\kms\ at 50 days after explosion);
\item  a small amount
of \ni\ produced in the explosion ($10^{-3}$ to $2 \times 10^{-2}$\,\msun,
which is 1--2 orders of magnitude less than in standard Type II SNe);
\item a low luminosity (pseudo-bolometric luminosity $L_{BVRI}$ ranges from $3.5 \times
10^{40}$ to $2.8 \times 10^{41}$~\ergs);
\item a low ejecta kinetic energy
(${\sim}\,0.1 \times 10^{51}$ to ${\sim}\,0.5 \times 10^{51}$~erg; \citealt{pumo_2p_17}).
\end{enumerate}

While Type II-P SNe represent  about 50--60\% of all core-collapse SNe
\citep{Smith2011, Graur2017}, the rate of low-luminosity Type II-P SNe
is currently estimated to be only 5\% of all Type II SNe \citep{Pastorello2004}.

Two distinct mass ranges have been studied for the RSG progenitors
of low-luminosity Type II-P SNe, corresponding to low-moderate mass massive
stars in a domain close to the lower-mass limit for core collapse
(\citealt{Chugai2000}, \citealt{Pignata2013}; \citealt{paper1}, hereafter L17; \citealt{pumo_2p_17}),
or to more massive stars in a domain closer to the limit where the progenitor dies
as a Wolf-Rayet instead of a RSG star \citep{Turatto1998, Zampieri2003}.
The progenitors of low-luminosity SNe II-P have, however, been constrained from
pre-explosion images.
The progenitor of SN\,2005cs has been identified as a RSG of spectral type K3--M4
with a main-sequence mass of 9--10\,\msun\ \citep{Maund2005,Li2006}.
For SN\,2008bk, the main-sequence mass of the RSG progenitor is estimated to be
8--13\,\msun\ \citep{VanDyk2012,Maund2014}.
For SN\,2009md, the main-sequence mass of the RSG progenitor is estimated to be
$8.5^{+6.5}_{-1.5}$\,\msun\ \citep{Fraser2011}, though \cite{Maund2015}
suggest that the disappearance of the progenitor cannot be confirmed.
The alternative of a high progenitor mass
is thus somewhat in tension with estimates from pre-explosion images.

\begin{figure*} 
\includegraphics[width=0.47\textwidth]{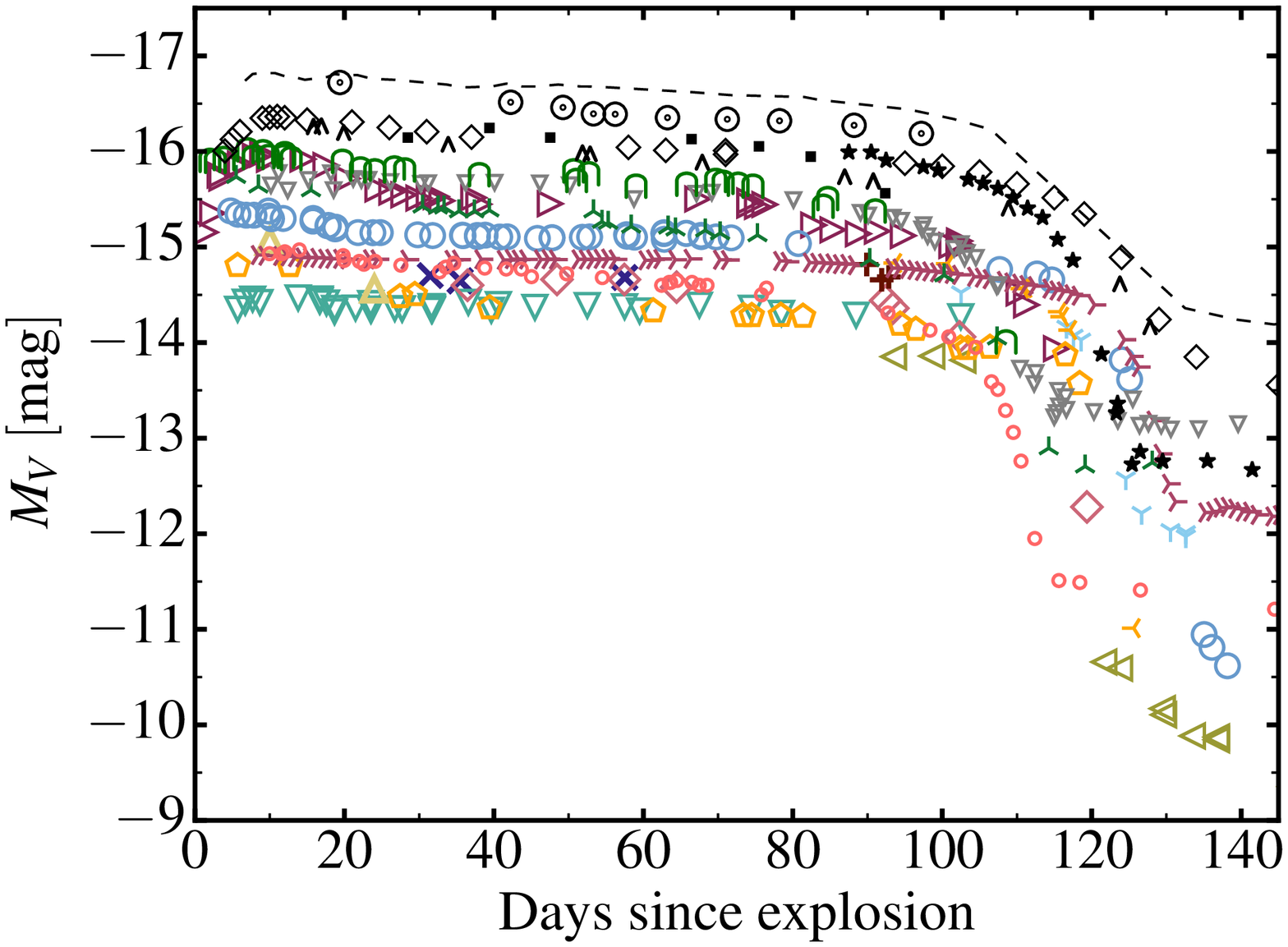}
\includegraphics[width=0.47\textwidth]{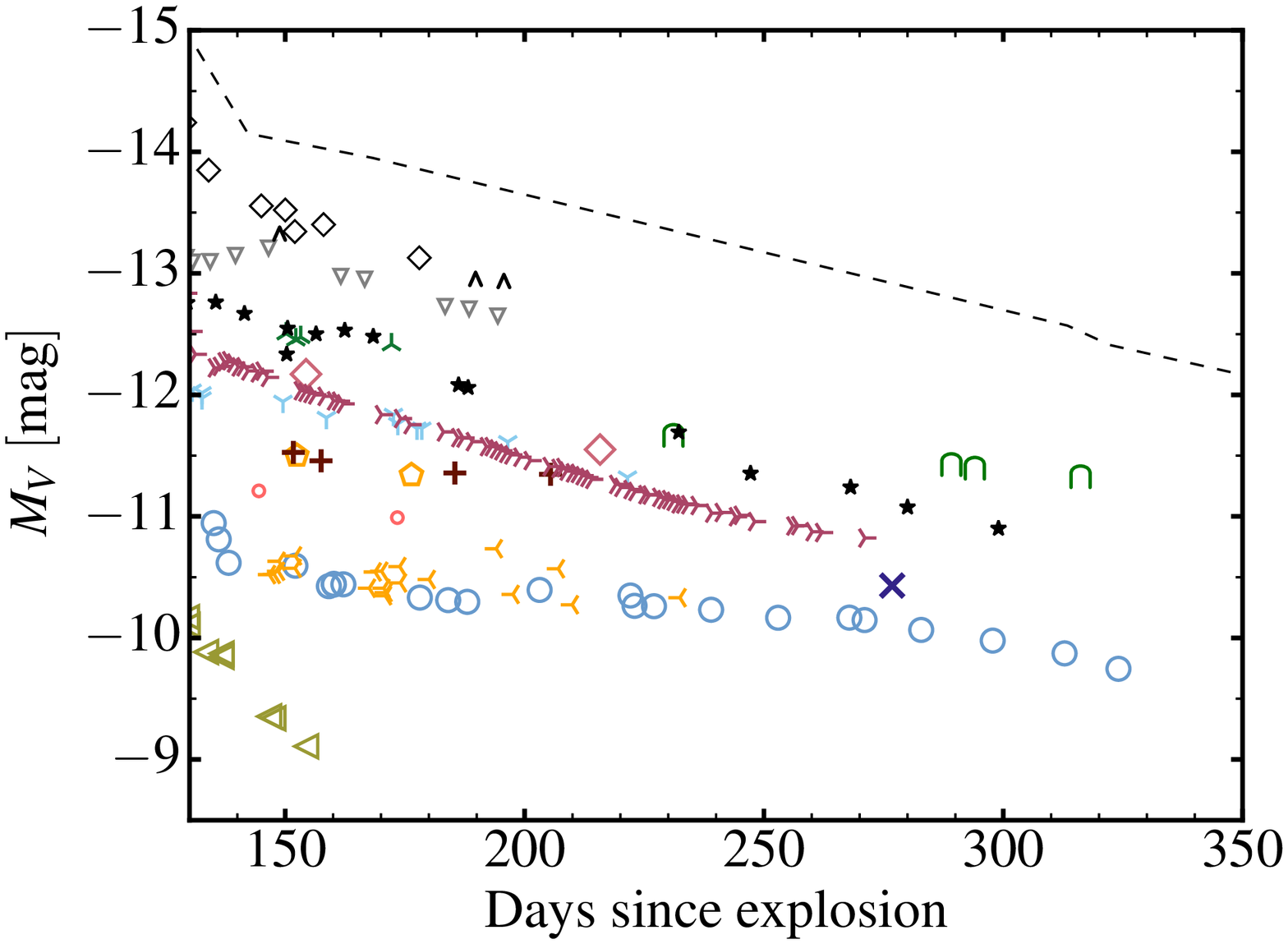}
\includegraphics[width=0.47\textwidth]{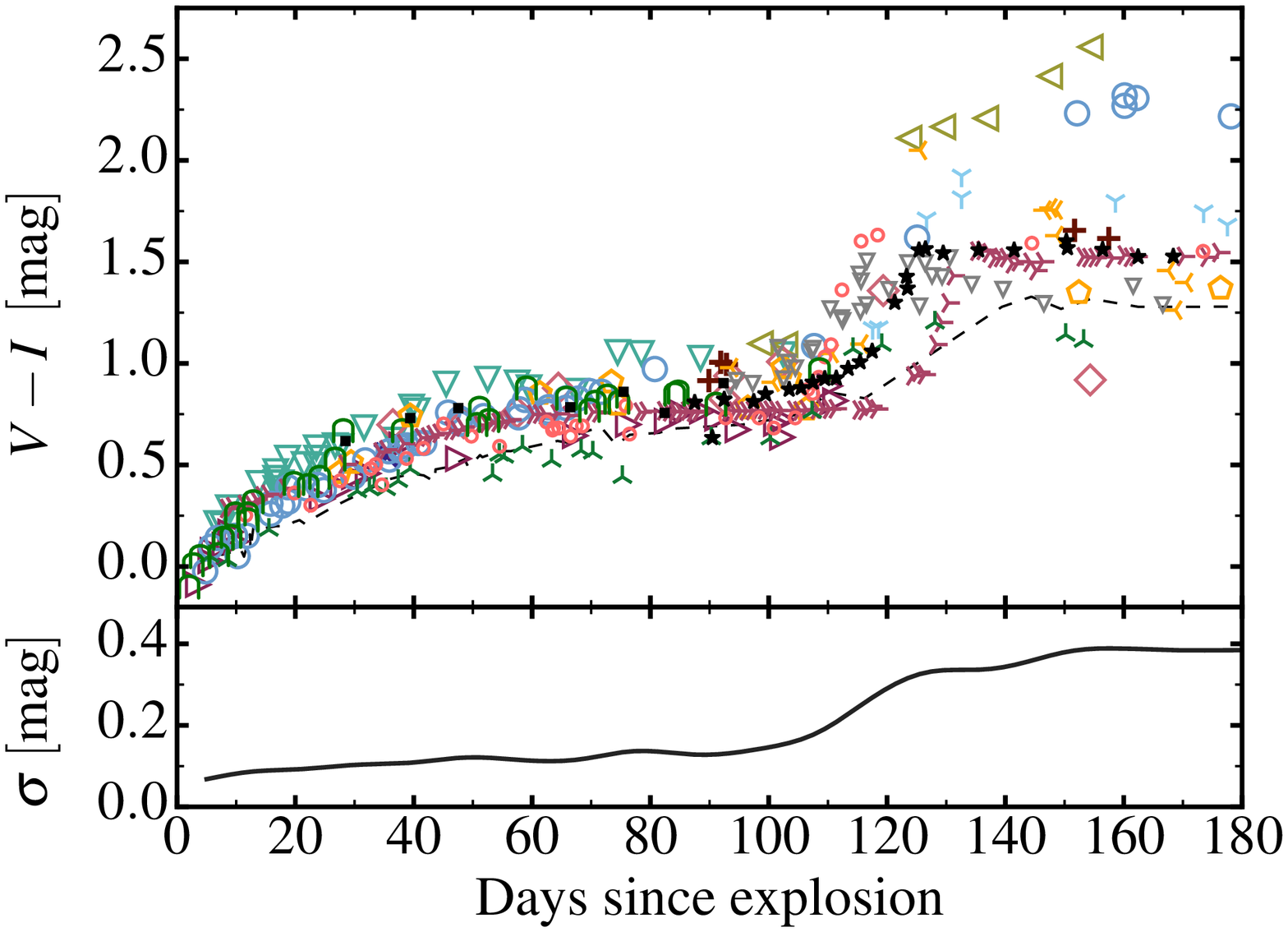}
\includegraphics[width=0.47\textwidth]{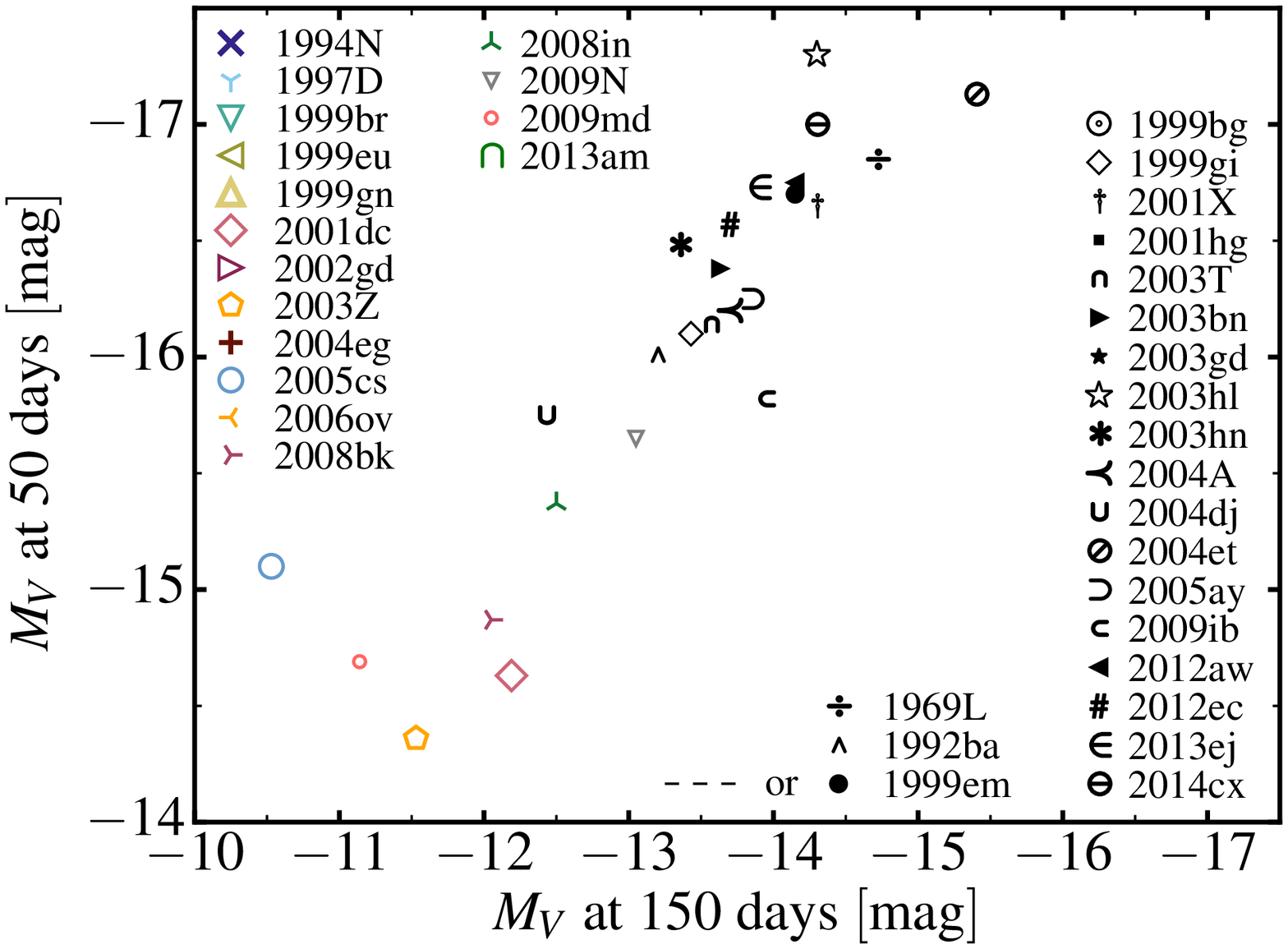}
\caption{Photometric properties of low-luminosity Type II SNe.
  We show the $V$-band light curves during the photospheric phase (top left)
  and the nebular phase (top right), the $V-I$ color evolution (together with
  the standard deviation at each time; bottom left), and the relation between
  the $V$-band brightness at 50 and 150\,d (bottom right;
  objects with no data around 50\,d or 150\,d are omitted). The time origin is
  the inferred time of explosion. All magnitudes have been corrected for
  distance and extinction. We overplot data for a small selection of SNe
  II-P (black symbols; not used for the standard deviation shown in the
  bottom left panel) that have intermediate plateau luminosities between the
  least luminous SN II-P (1999br) and the standard-luminosity SN II-P 1999em.
  There are no fast decliners in the current sample of low-luminosity SNe II.
  [See Section~\ref{analysis-obs-data} for discussion.]
   \label{fig_obs_quad}
   }
\end{figure*}

In our previous work (L17), we
performed a detailed study of the low-luminosity SN\,2008bk.
Guided by the inferred progenitor mass of 8--13\,\msun, based on pre-explosion photometry,
we explored a variety of progenitor massive stars dying with a final mass
of about 10\,\msun. In our set of seven models, the model (named X) 
that most closely matches the
observations is characterized at the time of collapse by a total mass of 9.88\,\msun,
log$(L/L_{\odot}) = 4.72$, and a surface radius of 502\,\rsun. The model ejecta kinetic energy
is $2.5\times10^{50}$\,erg, the ejecta mass is 8.29\,\msun, and the \ni\ mass
is $\sim$\,0.009\,\msun.
This model yields a fair match to the multi-band light curves and spectra of SN\,2008bk,
although it is somewhat too luminous and energetic.

In the present study, we consider all low-luminosity SNe II-P that we
could identify in the literature.
We model both low and high mass RSG star progenitors in order
to test whether, from the SN radiation alone, one can find evidence for the progenitor mass.
In practice, we compute low-energy explosions in high mass progenitors stemming
from a main sequence star of
25 and 27\,\msun. As discussed below, we find that the properties of the
SN radiation resulting from such massive progenitors have numerous points
of tension with observations, in contrast to its low mass counterpart
(a model for a 12\,\msun\ progenitor).

The paper is organized as follows.
In Section~\ref{analysis-obs-data}, we study the photometric and spectral properties of
our sample of low-luminosity Type II SNe and compare them to those of a standard SN II.
The observational data sources for this analysis are presented in Appendix~\ref{appendix:obs-data}.
In Section~\ref{modelling}, we present our modeling approach and initial conditions,
for the progenitor evolution, the star explosion, and the evolution of the ejecta
and radiation until late times.
In Section~\ref{results-lc} and Section~\ref{results-spectra}, we discuss our model results
for the multi-band light curves and spectra, respectively, and confront these results to observations.
In Section~\ref{other-work-comp}, we compare our results to other works.
Finally, we summarize our results in Section~\ref{conclusions}.

\section{Analysis of the observational data}
\label{analysis-obs-data}

The analysis presented in this section summarizes the properties
of low-luminosity SNe II-P.

The observations for these SNe have been presented in
\citet{Pastorello2004} (1994N, 1999br, 1999eu, 2001dc);
\citet{Spiro2014} (1999gn, 2002gd, 2003Z, 2004eg, 2006ov);
\citet{Benetti2001} (1997D);
\citet{Pastorello2006}, \citet{Dessart2008} and \citet{Pastorello2009} (2005cs);
\citet{Pignata2013} (2008bk);
\citet{Roy2011} (2008in);
\citet{Takats2014} (2009N);
\citet{Fraser2011} (2009md);
\citet{Gal-Yam2011} (2010id);
and \citet{Zhang2014} (2013am).

From the sample, we extract statistical properties that we later compare to
our model results for multi-band light curves and spectra. We consider the scatter
in their properties, and thus go beyond our previous study on a single event (SN\,2008bk; L17).
In Appendix~\ref{appendix:obs-data}, we present the sources of the observational
data (see also the information given in Tables~\ref{sn-data} and \ref{other-sn-data}).
In some cases, we revise the distance, the reddening and/or the recession velocity
published in the literature.

Although considered low-luminosity events, we exclude Type II-peculiar SNe like 1987A
since these events stem from the explosion of more compact stars like a blue supergiant
rather than a RSG --- their low luminosity stems primarily from the reduced progenitor
radius.

\subsection{Photometric properties}

We present $V$-band LCs for our sample in Fig.~\ref{fig_obs_quad} (top row
panels). All objects have a similar photometric evolution. The $V$-band
brightness appears as a plateau for the entire photospheric phase (i.e., before
the sudden drop that starts the nebular phase). Our sample occupies the faint
end of the SN II distribution. When combined with intermediate-luminosity events
(e.g., SN\,2011hg or 2003gd), there is no longer any luminosity/brightness gap
between the faintest event (SN\,1999br) and the standard
SN II-P 1999em.
Absolute $V$-band magnitudes during the
plateau phase lie between about $-14$ and about $-15.5$~mag (factor of $3-4$ in
luminosity). A standard SN II-P has $M_V{\sim}-16.7$~mag during the plateau
phase (Table~\ref{Anderson}). The continuous distribution of SNe II-P $V$-band
brightness suggests that there is also a continuous distribution in progenitor
and explosion properties rather than two separate classes.

In contrast to the scatter in plateau brightness, the plateau length is about
110$\pm$10\,d for the sample.
The plateau ends with an abrupt brightness
drop of 2.5--4.5\,mag over 10--20\,d and the SN enters the nebular stage. In
the absence of external energy sources, the nebular-phase luminosity is
powered by the radioactive decay of \,\co. While the standard Type II-P
SN\,1999em synthesized 0.04--0.06\,\msun\ of \ni\ \citep{Utrobin2007b,
Bersten2011}, the least luminous SN at nebular times in our sample,
SN\,1999eu, synthesized only about 0.001\,\msun\ \citep{Spiro2014}. Although
all our SNe show systematically lower \ni\ masses than SN\,1999em, they
exhibit much scatter in nebular $V$-band brightness, reflecting a sizeable
scatter in \ni\ masses (this dispersion may partly stem from differences in
bolometric correction or SN color).

\begin{figure*}
\includegraphics[width=0.99\textwidth]{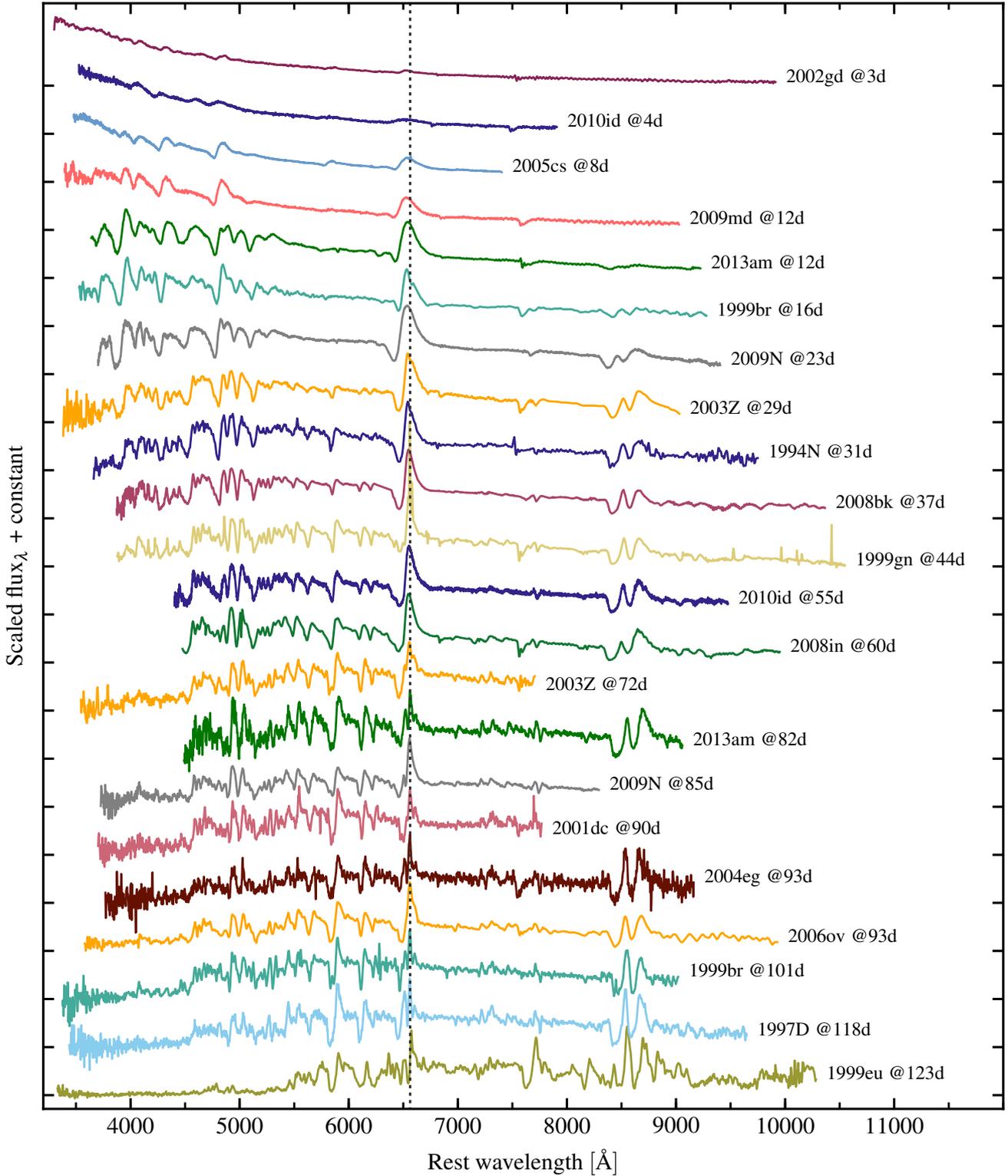}
\caption{Montage of spectra (corrected for redshift and reddening) showing the
  evolution of low-luminosity SNe II-P during the photospheric phase. Each of
  the 17 SNe in our sample is shown at least at one epoch. The
  ordinate ticks mark the zero flux level for each spectrum. The spectral
  evolution for these events is very generic, forming a smooth sequence
  towards redder optical colors and narrower spectral lines. The vertical line
  locates the H$\alpha$ rest wavelength and helps identifying any skewness in
  the line profile. For each spectrum, we indicate the phase with respect to
  the inferred time of explosion.
  \label{sne-montage}}
\end{figure*}

\citet{Anderson2014a} studied the $V$-band LCs for a sample of 116 Type II
SNe. To provide some quantitative comparison with this work, we show the mean
values and the standard deviation $\sigma$ for the   $M_{\rm max}$, $M_{\rm end}$,
$M_{\rm tail}$,  and $s_2$ in
Table~\ref{Anderson}. We adopt the same definition for these quantities (see
their Fig.~1):
$M_{\rm max}$ is the magnitude at the initial peak, if observed, otherwise it
corresponds to the first photometric point;
$M_{\rm end}$ is the $V-$band absolute magnitude measured 30\,d before
$t_{\rm PT}$, where
$t_{\rm PT}$ is the mid point of the transition from the plateau phase to the
radioactive tail;
$M_{\rm tail}$ is the $V-$band absolute magnitude 30\,d after $t_{\rm PT}$;
and $s_2$ is the decline rate in the $V-$band (given in magnitude per 100 days)
during the plateau phase.
Five SNe in our sample are also in the sample of \cite{Anderson2014a}.
The $V-$band decline rate is lower for our sample,
indicating that low-luminosity SNe show a $V-$band `plateau'; none of these events are
fast declining in the $V-$band during the photospheric phase.

\begin{table}
\caption{Photometric properties in the $V$-band for
  our sample of low-luminosity SNe II-P and for the
  larger sample of Type II SNe of \citet{Anderson2014a}.
  $M_{\rm max}$ is the magnitude at the initial peak if observed, otherwise first photometric
  point; $M_{\rm end}$ is the magnitude at the end of the plateau phase; $M_{\rm
  tail}$ is the magnitude at the beginning of the nebular phase, $s_2$
  is the decline rate during the `plateau' phase.
  See text for a more detailed description.
  \label{Anderson}}
\begin{tabular}{lcc} \hline
                       &Low-luminosity SNe II-P                     &SNe II \\ \hline
                       &$V$-band [mag]                              &$V$-band [mag]  \\
$\mean{M}_{\rm max}$   &$-15.29\,(\sigma=0.53, \:\,7\,{\rm SNe})$   &$-16.74\,(\sigma = 1.01, 68\,{\rm SNe})$ \\
$\mean{M}_{\rm end}$   &$-14.53\,(\sigma=0.50,    13\,{\rm SNe})$   &$-16.03\,(\sigma = 0.81, 69\,{\rm SNe})$ \\
$\mean{M}_{\rm tail}$  &$-11.65\,(\sigma=0.81,    11\,{\rm SNe})$   &$-13.68\,(\sigma = 0.83, 30\,{\rm SNe})$ \\ \hline
                       &$V$-band [mag per 100 days]                 &$V$-band [mag per 100 days] \\
$\mean{s_2}$           &0.25 ($\sigma=0.08$, 10  SNe)               &1.27 ($\sigma = 0.93$, 113 SNe)  \\ \hline
\end{tabular}
\end{table}

We present the color evolution $V-I$ for our set of SNe in the bottom-left panel of
Fig.~\ref{fig_obs_quad}. The scatter is small during the photospheric phase and increases during the
nebular phase. All the SNe from our sample evolve in a very similar way irrespective of the
plateau brightness. At early times after explosion, the color continuously reddens until the SN enters
the recombination phase, during which its optical color is roughly constant. At the end of the plateau
phase, the ejecta become transparent and the value of $V-I$ rises rapidly, revealing also a
large scatter. At nebular times, the flux is dominated by lines, whose wavelength distribution and
relative strength control the color. If one includes moderate- and standard-luminosity SNe, such as
1992ba, 1999bg, 1999gi, 2001hg, 2002ca, 2003gd, 2005ay, or 2012ec, the dispersion in $V-I$ color is
not affected.

In L17, we showed how the treatment and magnitude of \ni\ mixing has a strong impact on the nebular
phase color. The observed scatter of data points suggests that the mixing process may differ amongst
low-luminosity SNe II-P. In the bottom-right panel of Fig.~\ref{fig_obs_quad} we compare the plateau
and nebular phase luminosities. Although the scatter is significant, there is a strong correlation. It
indirectly connects the explosion energy and the mass of ejected \ni.
More energetic explosions tend to produce more \ni\ \citep{Ugliano2012,muller_56ni_17}.

\subsection{Spectral properties}

\begin{figure}
\includegraphics[width=0.47\textwidth]{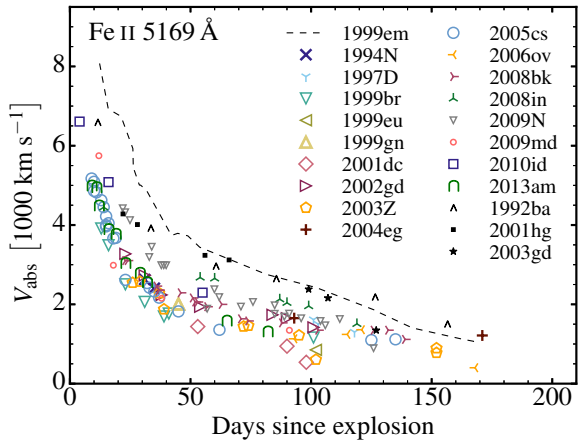}
\caption{Doppler velocity at maximum absorption in Fe\two\,5169\,\AA\  for our
  sample of low-luminosity SNe II-P and other, more energetic, Type II-P SNe,
  including 1999em, 1992ba, 2001hg and 2003gd. Relative to SN\,1999em,
  low-luminosity SNe II-P  have systematically lower expansion rates. This
  suggests they not only have a low luminosity but also a low ejecta kinetic
  energy.
  \label{vel-sn} }
\end{figure}

In Fig.~\ref{sne-montage}, we show the spectral evolution for our set of SNe.
All of them follow the same pattern and show very similar features in the
spectra. Low-luminosity SNe II-P systematically exhibit much narrower lines
than standard SNe II-P. Figure~\ref{vel-sn} shows that the Doppler velocity at
maximum absorption in Fe\two\,5169\,\AA\ is few tens of percent smaller than
for standard SNe II-P, corresponding to a slower ejecta expansion rate.
Consequently, their spectra suffer much less line overlap and individual lines
are more easily identified.

\citet{DH2005a} discussed the optical-depth effect at the origin of the blue-shifted emission peaks
of P-Cygni profiles in SNe II (see also \citealt{Anderson2014b} for a discussion
on alternative interpretations for the origin of this feature).
This blue shift is observed in all Type II SN
spectra irrespective of $V$-band decline rate \citep{Anderson2014b}. In
H$\alpha$, the peak blue shift is greatest at early times, decreases through
the photospheric phase, and vanishes as the SN becomes nebular. The spectra
for our sample of low-luminosity SNe show the same behavior.

The evolution of the spectral morphology of standard Type II-P SNe has been
discussed numerous times, both from observational data and tailored models
(see, e.g., \citealt{Leonard2002b, DH2011}). This evolution is the same for
low-luminosity  SNe II-P. At very early times ({$\lesssim$}5 days since
explosion), the spectra have a color temperature greater than $10^{4}$\,K, are very
blue, and show weak lines of H\one, He\one--\!\two, and from neutral (and more
rarely once ionized) species of C, N, or O. Metal lines (in particular Ti\two,
Fe\two), which eventually cause line blanketing, start to develop as the
photospheric layers recombine, which takes place after about two weeks. This
is accompanied by the strengthening of the NIR Ca\two\ triplet at about
8500\,\AA, Na\one\,D, lines of Sc\two\ and Ba\two. Many of these lines remain
strong until the nebular phase, in part because they are tied to low lying
levels which can be more easily excited (thermally or non-thermally).

\begin{figure}
\includegraphics[width=0.47\textwidth]{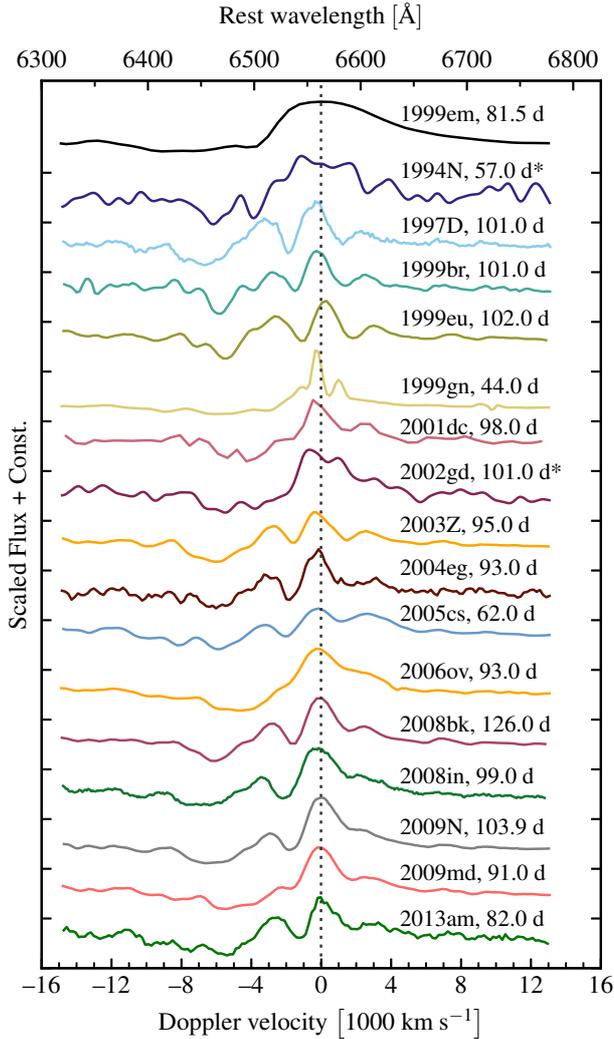}
\caption{
Spectral montage of the H$\alpha$ region for our sample of low-luminosity SNe II-P
at $\sim$\,100\,d after explosion.The complex structure in the
H$\alpha$ profile is better seen at that time. If no data is available at this epoch, we use
the closest observation. The ordinate ticks mark the zero flux level for each spectrum.
The top axis shows the rest wavelength. Spectra marked with an asterisk have been
smoothed to reduce the noise level.
\label{halpha-ba2}}
\end{figure}

The H$\alpha$ profile becomes structured at the end of the plateau phase in
low-luminosity SNe II-P (Fig.~\ref{halpha-ba2}; \citealt{Pastorello2004};
\citealt{Roy2011,Spiro2014}). Standard Type II SNe rarely show a complex
H$\alpha$ profile because their higher expansion rates cause a stronger
Doppler broadening and line overlap. In L17, we computed a radiative-transfer
model that suggests that Ba\two\,6496.9\,\AA\ causes the structure seen in
H$\alpha$ in low-luminosity SNe II-P --- overlap with Sc\two\ causes additional
structure in the red part of H$\alpha$ (See Fig.~9 in L17). Ejecta asymmetry
may also contribute, but it does not appear essential.

\begin{figure}
\includegraphics[width=0.47\textwidth]{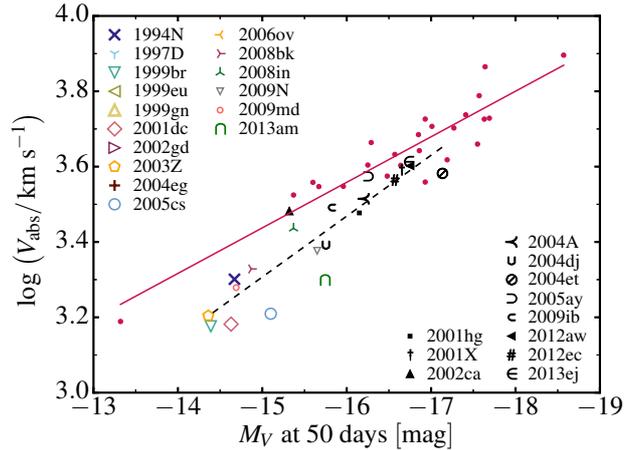}
\caption{Variation of the Doppler velocity at maximum absorption in Fe\two\,5169\,\AA\
  with the intrinsic $V$-band brightness 50\,d after explosion for low-luminosity SNe II-P
  (colored symbols; we exclude SN\,2010id since its photometry seems anomalous,
  see Appendix~\ref{2010id}) and standard-luminosity SNe II-P (black symbols).
  When necessary, photometric/velocity measurements have been
  interpolated to a post-explosion epoch of 50\,d.
  Small red filled circles correspond to the Type II SN sample from \citet{Hamuy2003},
  not described in this text, and the solid red line is a fit to these data points.
  The dashed black line is a fit to all the SNe listed in the legend.
  The fitted lines are of the form $\log V_{\rm abs} = a + b M_V$, where $V_{\rm abs}$
  is in~\kms.
  For the low-luminosity sample, we find $a= 0.88$ and $b=-0.16$, and for
  the sample from \citet{Hamuy2003}, we find $a=1.62$ and $b=-0.12$.
   \label{mv50-vel-fe-50} }
\end{figure}

\subsection{Visual brightness versus expansion rate}

Observations indicate that intrinsically brighter Type II-P SNe have higher
photospheric velocities half-way through the photospheric phase \citep{Hamuy2003}.
Numerical simulations of RSG star
explosions naturally predict such a correlation \citep{Popov1993,KW2009,DLW2010b}.
Using the information from Fig.~\ref{vel-sn}, we extend the sample
of \citet{Hamuy2003} and include our measurements for low-luminosity SNe II-P
(Fig.~\ref{mv50-vel-fe-50}).

The correlation identified by \citet{Hamuy2003} extends to low-luminosity SNe II-P,
although the slope is altered slightly and the scatter is significantly larger at
the low brightness end. While uncertainties in distance might play a role,
this scatter may reflect differences in progenitor/explosion properties. For example,
for the same explosion energy and progenitor H-rich envelope mass, a larger
radius would produce a brighter plateau and a lower expansion rate \citep{Dessart2013}.

\section{Modelling} \label{modelling}

\begin{figure}
\includegraphics[width=0.48\textwidth]{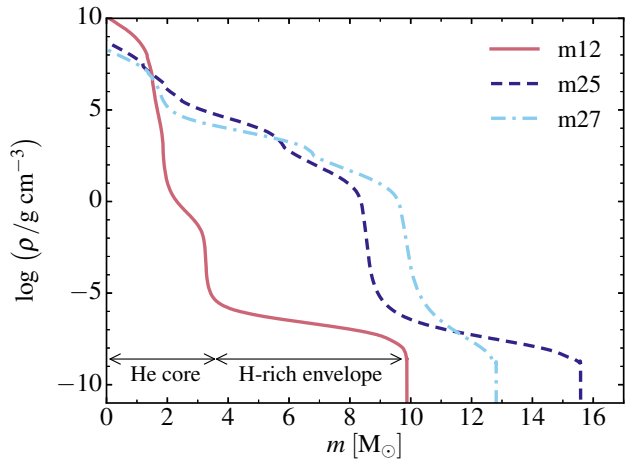}
\caption{Density versus lagrangian mass for our set of pre-SN models, evolved
  with~\mesa. The extended H-rich envelope is the external
  region with $\rho < 10^{-5}\,{\rm g\,cm^{-3}}$ and the region below is the He
  core (these two regions are indicated for model m12).
  The other large density jump that appears deeper in the star corresponds to the transition
  to the CO core.
  \label{mesa-rho} }
\end{figure}

The numerical approach followed in this work is identical to the one presented
in \cite{Dessart2013}. It consists of simulations for the progenitor star from
main sequence to core collapse with \mesa\
\citep{Paxton2011,Paxton2013,Paxton2015}, its subsequent explosion with the
radiation-hydrodynamics code \v1d\ \citep{Livne1993, DLW2010a,DLW2010b}, and
the evolution until late times with the time-dependent radiative-transfer code
\cmfgen\ \citep{HM1998,DH2005a,DH2008,HD2012,Dessart2013}. We briefly review
each step in the forthcoming sections.

By modeling the observed SN II-P LCs and spectra, we aim to constrain the
ejecta and progenitor properties. Doppler-broadened spectral lines can be used
to infer the expansion rate. The Type II SN plateau duration correlates with
the progenitor radius and H-rich envelope mass, as well as the explosion
energy \citep{Arnett1980, Litvinova1983, Litvinova1985, Popov1993, Young2004,
KW2009, Dessart2013}. The SN color evolution correlates with the progenitor
radius \citep{Dessart2013}. In this paper, we investigate how the different
properties of low- and high-mass RSG stars impact the SN II-P observables. For
example, the He-core mass increases with main sequence mass
\citep{Woosley2002}. This then impacts the stellar luminosity, which affects
both the envelope mass (through the effect of mass loss) and the envelope
radius (through the constraint of radiative diffusion).

\subsection{Pre-SN evolution with \mesa}\label{mesa}
\label{sect_mesa}

Using \mesa, we evolve three massive star models with an initial mass
of 12, 25 and 27\,\msun\ (named m12, m25 and m27) from the main sequence
until core collapse. We do not consider binary star evolution,
which can alter the relation between the mass on the zero-age-main-sequence
and the mass of the SN progenitor at the time of explosion.
We use the same parameters as in L17. Model m12 is the same
as model X from L17. For the new models m25 and m27, we use $Z=0.0162$ rather than 0.02.
Models m12, m25 and m27 end their lives with final masses of
9.9, 15.6 and 12.8\,\msun, and surface radii of 502, 872 and 643\,\rsun.
A summary of model properties is given in Table~\ref{prog-ej-prop}.
We adopt the same prescriptions for the convection, mass loss, core overshooting etc.
in all models. While the processes controlling massive star evolution are not accurately
known or described, the trends that emerge from our study should be robust.
Our progenitor models are in good agreement with those of \citet{Woosley2002}.

Figure~\ref{mesa-rho} shows the density profile for the three models at the onset
of core collapse. In this figure, the extended H-rich envelope corresponds to the region with
$\rho < 10^{-5}\,{\rm g\,cm^{-3}}$, which is exterior to the He core
(its outer edge coincides with the large density jump at the base of the H-rich envelope).
We take the location where the H mass fraction suddenly
drops from its surface value (the whole convective envelope is homogeneous) as
the interface between the H-rich envelope and the He core.
With this definition, the H-rich envelope mass is 6.6--7.0\,\msun\ for the m12 and m25 models,
but it is only 3\,\msun\ in the m27 model due to the greater
mass lost during the RSG phase. In single stars, the H-rich envelope mass depends
on the adopted wind mass loss rate, which is uncertain, but the trend is robust. In particular,
while models m12 and m25 have a similar H-rich envelope mass, they
have a very different He core mass of about 3.3 and 8.6\,\msun.
The core/envelope mass ratio in model m12 ($3.3/6.6$) is reversed in model m27
($9.8/3.0$).
This reversal in core/envelope mass ratio is fundamental for understanding 
the difference in SN properties resulting from the explosion of low- and high-mass 
RSG stars.

\begin{table}
\caption{Mean velocities for the whole ejecta or for the H-rich ejecta only, and
velocity at the base of the H-rich layer for our models m12, m25, and m27.
[See text for details.]
\label{mean-vel}}
\centering
  \begin{tabular}{lccc} \hline
        &$\mean{V}_{\rm ej}$  &$\mean{V}_{\rm H}$  &$V_{\rm H,min}$ \\ \hline
        &[\kms]                 &[\kms]              &[\kms] \\ \hline
    m12 &1550                   &1800                & 800   \\
    m25 &1900                   &2200                &1320   \\
    m27 &2200                   &3150                &2200   \\ \hline
  \end{tabular}
\end{table}

\begin{figure*}
\includegraphics[width=0.31\textwidth]{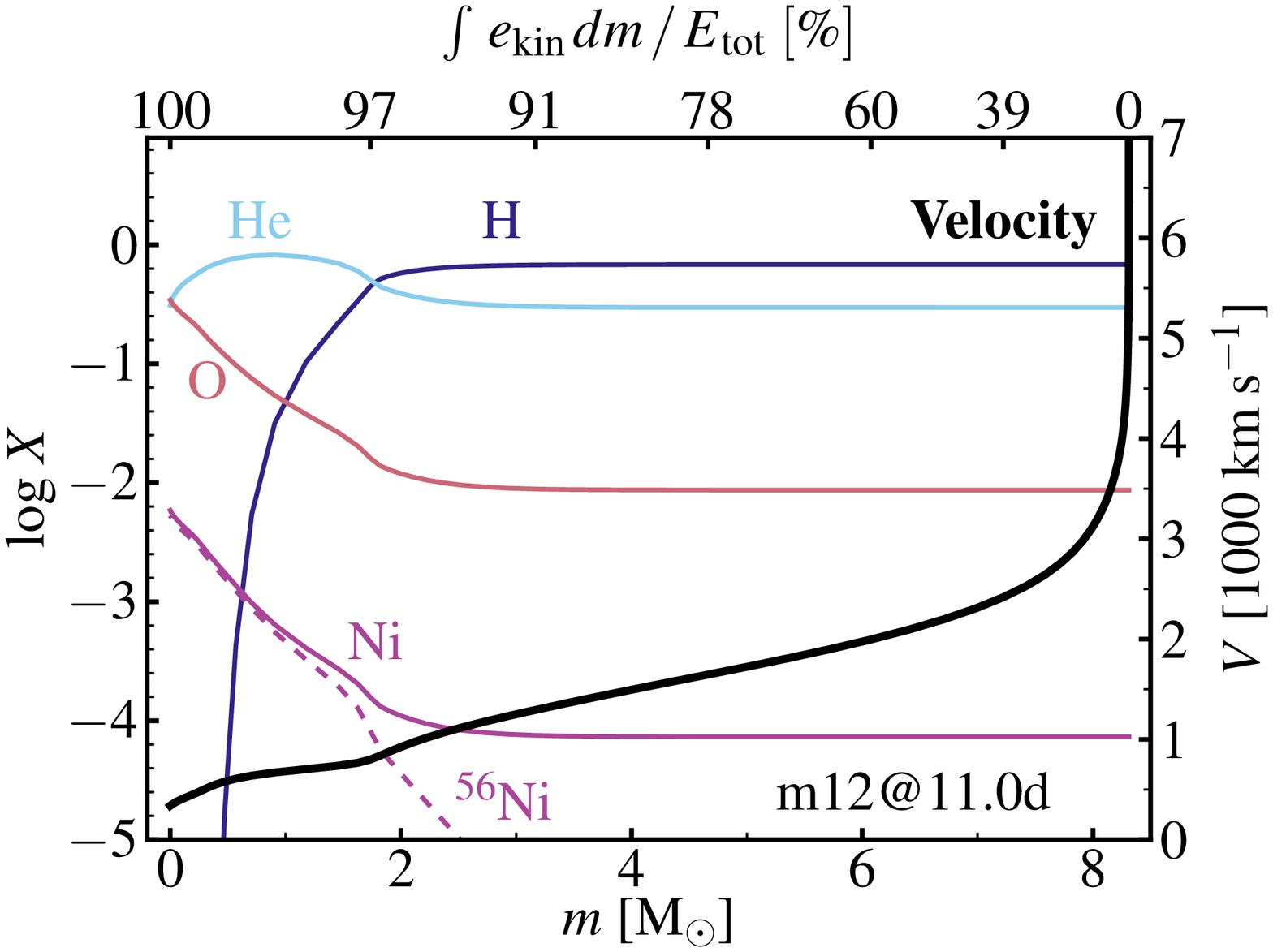}
\hspace{3mm}
\includegraphics[width=0.31\textwidth]{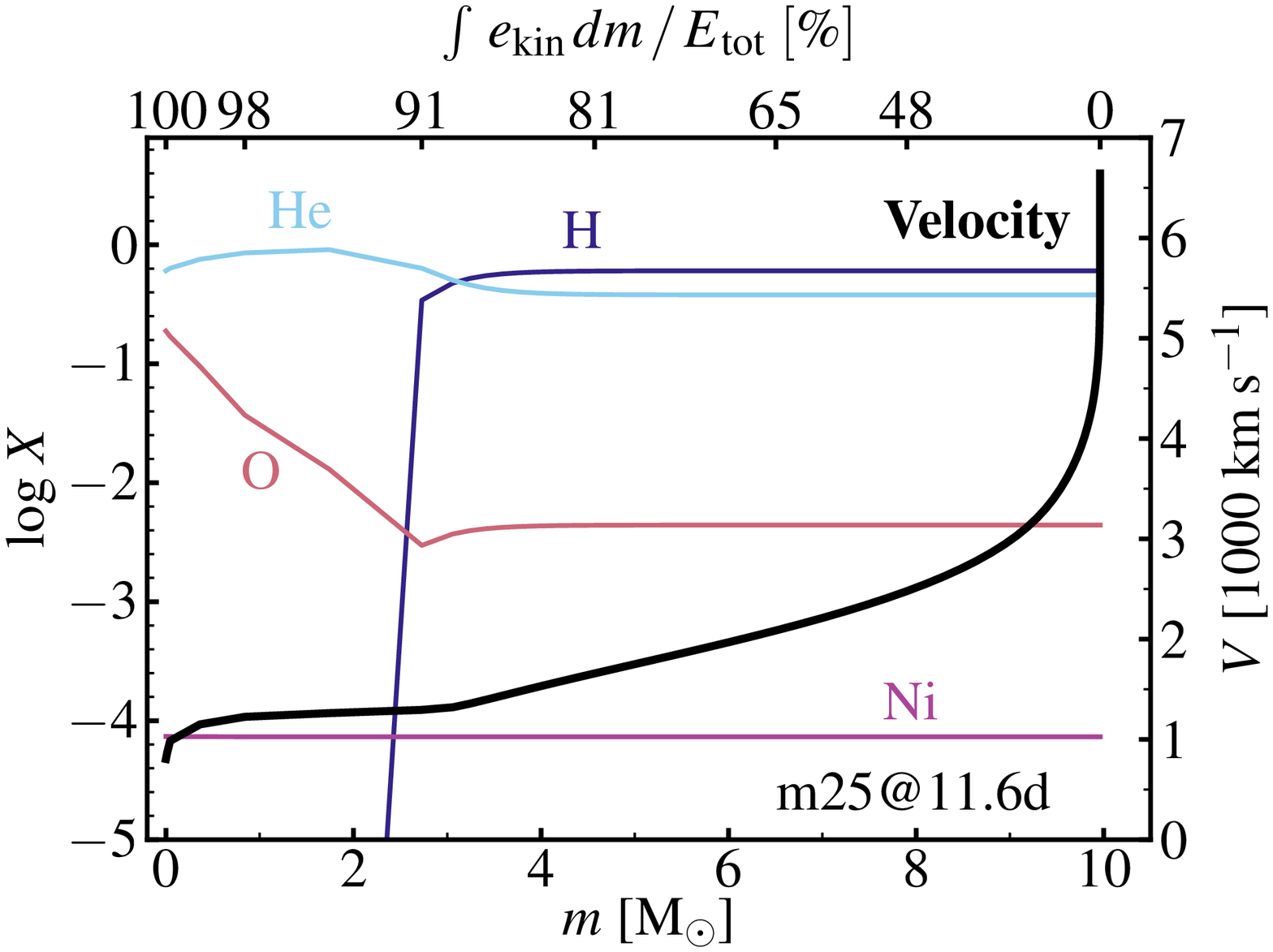}
\hspace{3mm}
\includegraphics[width=0.31\textwidth]{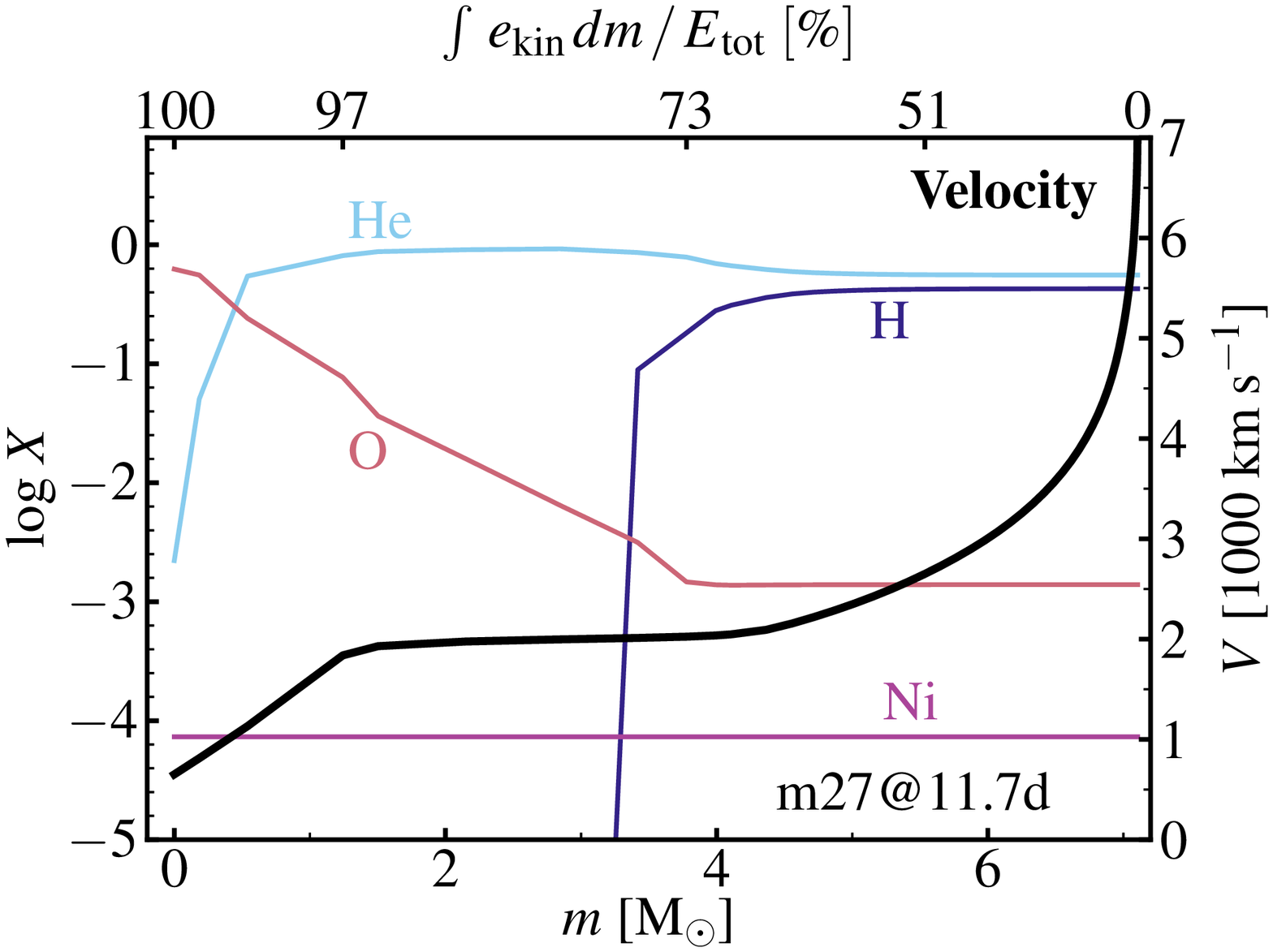}
\caption{Variation of the mass fraction for H, He, O, and Ni (the dotted line corresponds
to \ni\ in model m12; left panel) with ejecta lagrangian
  mass for models m12, m25, and m27.
  We overplot the velocity (thick line; right axis) and indicate the depth variation of the
  fractional inward-integrated kinetic energy (top axis).
  [See Section~\ref{sect_mesa} for discussion.]
\label{mf-mass} }
\end{figure*}

\subsection{Piston-driven explosion with \v1d}
\label{sect_v1d}

At the onset of core collapse, the \mesa\ models are exploded with the radiation
hydrodynamics code \v1d\  by moving a piston at $\sim$\,10,000\,\kms.
The mass cut for the piston location is where the progenitor entropy rises
outward from the centre to 4~$k_{\rm B}$\,baryon$^{-1}$ (see, e.g., \citealt{Ugliano2012}).
This location is at a lagrangian mass coordinate of 1.51, 1.93, and 1.78\,\msun\
in models m12, m25, and m27.

To produce our models of low-luminosity SNe II-P, the piston is kept at 10000\,\kms\ until
the deposited energy exceeds the binding energy of the material above the
piston by an amount $E_{\rm kin}$, where $E_{\rm kin}$ is the ejecta kinetic
energy at infinity. The binding energy of the material above the piston mass cut is
1.14, 7.47, and $5.18 \times 10^{50}$\,erg in models m12, m25, and m27,
while we aim to produce an ejecta
with $E_{\rm kin} \sim 2 \times 10^{50}$\,erg (which is the $E_{\rm kin}$
inferred for the prototypical low-luminosity SN\,2008bk; \citealt{Pignata2013};
L17).  This value of $E_{\rm kin}$ is therefore of the order or smaller
than the binding energy above the piston.
To prevent the hydrodynamical simulation from going on hold because of
a Courant-time limitation, we set a minimum piston velocity of 100\,\kms\ (rather than zero) 
in these simulations.
This prevents the growth of a hot and dense shell with negative velocities above
the inner boundary. We consider as fallback material any material moving slower
than 150$-$200\,\kms\ at 10$^6$\,s after the piston trigger. With this assumption, our
weak explosions produce significant fallback material in the inner
layers, much more than in standard SNe II-P where the ejecta  kinetic energy exceeds
the binding energy of the material to expel (see., e.g., \citealt{DLW2010a}).
In models m12, m25, and m27, the fallback mass
(i.e., envelope material moving within a factor of 1.5$-$2 of the asymptotic piston velocity) 
is  0.08, 3.69, and 4.0\,\msun.
In m12, this means that most of the Si-rich layer falls back, while in models
m25 and m27, it is the entire CO core that falls back, leading to the formation
of a 5--6\,\msun\ black hole.
In models with fallback, it is not straightforward to predict the kinetic energy of the
ejected material. Here, our ejecta have a kinetic energy of
2.5, 4.2, and 4.2$\times$10$^{50}$\,erg in models m12, m25, and m27.
While about 0.009\,\msun\ of \ni\ is expelled in model m12, the strong
fallback in models m25 and m27 prevents any ejection of \ni.
We have not tried to prevent this by additionally enhancing the mixing
(we use the same mixing in models m12, m25, and m27; see L17 for details).
Lacking a decay power source, the ejecta of models m25 and m27 produce a negligible
luminosity at nebular times. Hence, most of our discussion will be focused
on the photospheric phase, when the photosphere is located in the H-rich
layers of the progenitor star.

We show the composition profile for the ejecta for our models in mass and
velocity space in Fig.~\ref{mf-mass}. For model m12, about 50\% of the total
ejecta kinetic energy is contained in the outer 2\,\msun\ of the ejecta, and
only a few percent in the former He core (below 2000\,\kms). For models m25 and
m27, the former He core contains $\sim$\,9\% and $\sim$\,25\% of the total ejecta
kinetic energy, respectively.  The mass-weighted mean velocity of the whole ejecta
($\mean{V}_{\rm ej}$) and of the H-rich material ($\mean{V}_{\rm H}$) are
given in Table~\ref{mean-vel},  together with the velocity at the junction between
H-poor/H-rich layers (corresponding to the former core/envelope transition;
$V_{\rm H,min}$). This value correlates with the minimum width of H$\alpha$
in the Type II SN spectrum \citep{DLW2010b}.

Because of variations in $E_{\rm kin}$ and/or $M_{\rm ej}$
and differences in the chemical/mass stratification, the
$\mean{V}_{\rm H}$ and $V_{\rm H,min}$ vary significantly
between models. These variations will have a clear impact on
the resulting SN observables, which are discussed in
Sections~\ref{results-lc} and \ref{results-spectra}.
The value $V_{\rm H,min}$ is, however, uncertain because it is not clear
how much and how deep H will be mixed inwards.
\citet{wongwathanarat_15_3d} have demonstrated
that in a standard energy explosion of a 15\,\msun\ progenitor, H may be mixed
all the way to the innermost layers. No simulation has provided reliable
constraints for H mixing in a low energy explosion of a higher mass star, in which the
He core mass is much larger (and may exceed the H-rich envelope mass) and in
which strong fallback occurs.
We note that strong inward mixing of H is not guaranteed. Type IIb SNe are a notorious example
since they show broad H$\alpha$ typically for 1-2 weeks. H$\alpha$ is
absent in the nebular-phase spectra of SN\,2011dh
\citep{jerkstrand_15_iib}. H mixing is perhaps facilitated
in progenitors with a small He core mass and a massive H-rich envelope,
hence lower mass stars on the main sequence.

It is interesting to compare our model set to the simulations of \citet{sukhbold_16},
in particular those produced using a light bulb mimicking a neutrino-driven explosion,
nicknamed P-HOTB.
Model m12 properties correspond
closely to the 9$-$12\,\msun\ models of \citet{sukhbold_16} exploded with the Z9.6 engine,
which systematically yield low energy explosions, a low/moderate \iso{56}Ni mass,
and a neutron star remnant.
The 25\,\msun\ progenitor models of \citet{sukhbold_16} all explode with a 10$^{51}$\,erg
ejecta kinetic energy with a large \iso{56}Ni mass, in contrast with our model m25.
However, the model 27.3 of \citet{sukhbold_16}, exploded with P-HOTB using the W18
calibration yields an ejecta devoid of \iso{56}Ni, with a kinetic energy of $4.1 \times 10^{50}$\,erg,
and leaves behind a black hole of 6.24\,\msun. These properties are similar to those of model m27.
Hence, our models based on piston-driven explosions have counterparts in the more physically
consistent explosion models of \citet{sukhbold_16}, although the latter depend significantly
on the way the explosion engine is calibrated (engines N20 and W18 can yield drastically
different outcomes for the same progenitor).

\subsection{Radiative-transfer modeling with CMFGEN}
\label{sect_cmfgen}

At a post-explosion time of $\sim$\,11\,d, the ejecta are close to being in
homologous expansion. We then remap the \v1d\ ejecta structure and composition
into the non-Local-Thermodynamic-Equilibrium (nLTE)
time-dependent radiative transfer code \cmfgen\  and model the
subsequent evolution of the gas and the radiation until nebular times. The
code computes the gas and radiation properties by solving iteratively the
statistical equilibrium equations, the gas-energy equation and the first two
moments of the radiative transfer equation --- time-dependent terms are accounted
for in all equations. Our numerical setup (numerical grid, atomic data, model atoms)
is identical to L17. We present the results of the
radiative-transfer modeling in the next section.

\begin{figure}
\includegraphics[width=0.48\textwidth]{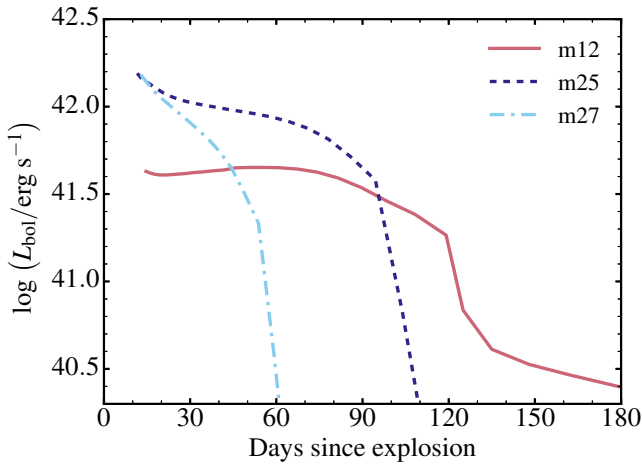}
\caption{\cmfgen\ bolometric light curves for models m12, m25, and m27.
In this sequence, the light curve evolves from a II-P to a II-L morphology,
with a greater luminosity at earlier times and an earlier transition
to the nebular phase.
\label{fig_lbol}}
\end{figure}

SN\,2008bk follows closely the average brightness, expansion rate, and
color during the photospheric and nebular phases of our sample of
low-luminosity SNe II-P (Figs.~\ref{fig_obs_quad}--\ref{vel-sn}).
SN\,2008bk can therefore be used as a template for this class of events
when confronting models to observations of low-luminosity SNe II-P.

\begin{table*}
  \caption{Summary of progenitor and ejecta properties for our models m12, m25,
    and m27. The left half of the table gives the initial mass $M_{\rm i}$ and
    pre-SN properties ($M_{\rm f}$, $R_\star$, $T_{\rm eff}$, $L_\star$, H-rich
    envelope mass, He-core mass, binding energy above the piston mass
    cut). The right half gives some properties of the corresponding ejecta,
    i.e., the ejecta mass, the remnant mass, the total yields for H, He, O, the
    amount of \ni\ synthesized in the explosion and the asymptotic ejecta
    kinetic energy. Numbers in parentheses correspond to powers of ten.
    \label{prog-ej-prop} }
  \begin{tabular}{lccccccccccccccc} \hline
    Model &$M_{\rm i}$ &$M_{\rm f}$ &$R_\star$ &$T_{\rm eff}$ &$L_\star$ &H-rich  &He-core &$E_{\rm b}$ &$M_{\rm ej}$ &$M_{\rm remnant}$ &H       &He      &O       &\ni        &$E_{\rm kin}$ \\ \hline
          &[\msun]     &[\msun]     &[\rsun]   &[K]           &[\lsun]   &[\msun] &[\msun] &[B]         &[\msun]      &[\msun]           &[\msun] &[\msun] &[\msun] &[\msun]    &[B]           \\ \hline
    m12   &12          &9.9         &502       &3906          &52733     &6.6     &3.3     &0.11        &8.29         &1.59              &4.54    &3.24    &0.22    &8.57($-3$) &0.25          \\
    m25   &25          &15.6        &872       &4299          &233050    &7.0     &8.6     &0.75        &9.98         &5.62              &4.34    &5.17    &0.13    &0          &0.42          \\
    m27   &27          &12.8        &643       &5227          &276761    &3.0     &9.8     &0.52        &7.02         &5.78              &1.37    &4.72    &0.4     &0          &0.42          \\ \hline
  \end{tabular}
\end{table*}

\begin{table*}
  \caption{Sample of results for our set of simulations. $\Delta t_{\tau>1}$ gives the post-explosion time when the ejecta turns
  optically thin to electron scattering. We then quote the values at 15 and 50\,d after explosion of the bolometric luminosity,
  the $V$-band magnitude, the $U-V$ color, the photospheric velocity, and the Doppler velocity at maximum absorption
  in H$\alpha$. Numbers in parentheses correspond to powers of ten.
    \label{sample} }
  \begin{tabular}{lccccccccccc} \hline
  Model
    &$\Delta t_{\tau>1}$ [d]
      &\multicolumn{2}{c}{$L_{\rm bol}$ [\ergs]}
        &\multicolumn{2}{c}{$M_V$ [mag]}
          &\multicolumn{2}{c}{$U-V$ [mag]}
            &\multicolumn{2}{c}{$V_{\rm phot}$ [\kms]}
              &\multicolumn{2}{c}{$V$(H$\alpha$) [\kms]} \\ \hline
          &      &(15d)      &(50d)    &(15d)    &(50d)    &(15d)    &(50d)  &(15d)   &(50d)  &(15d)  &(50d)    \\ \hline
  m12     &131   &0.45(42)   &4.8(41)  &--15.49  &--15.61  &--0.06   &2.01   &4833    &2401   &5290   &3910     \\
  m25     &112   &1.38(42)   &9.3(41)  &--16.22  &--16.50  &--0.87   &1.46   &5065    &3371   &5724   &4213     \\
  m27     &64    &1.37(42)   &3.0(41)  &--16.38  &--15.28  &--0.72   &2.95   &6251    &2498   &6706   &6062     \\ \hline
  \end{tabular}
\end{table*}

\begin{figure*}
\includegraphics[width=0.48\textwidth]{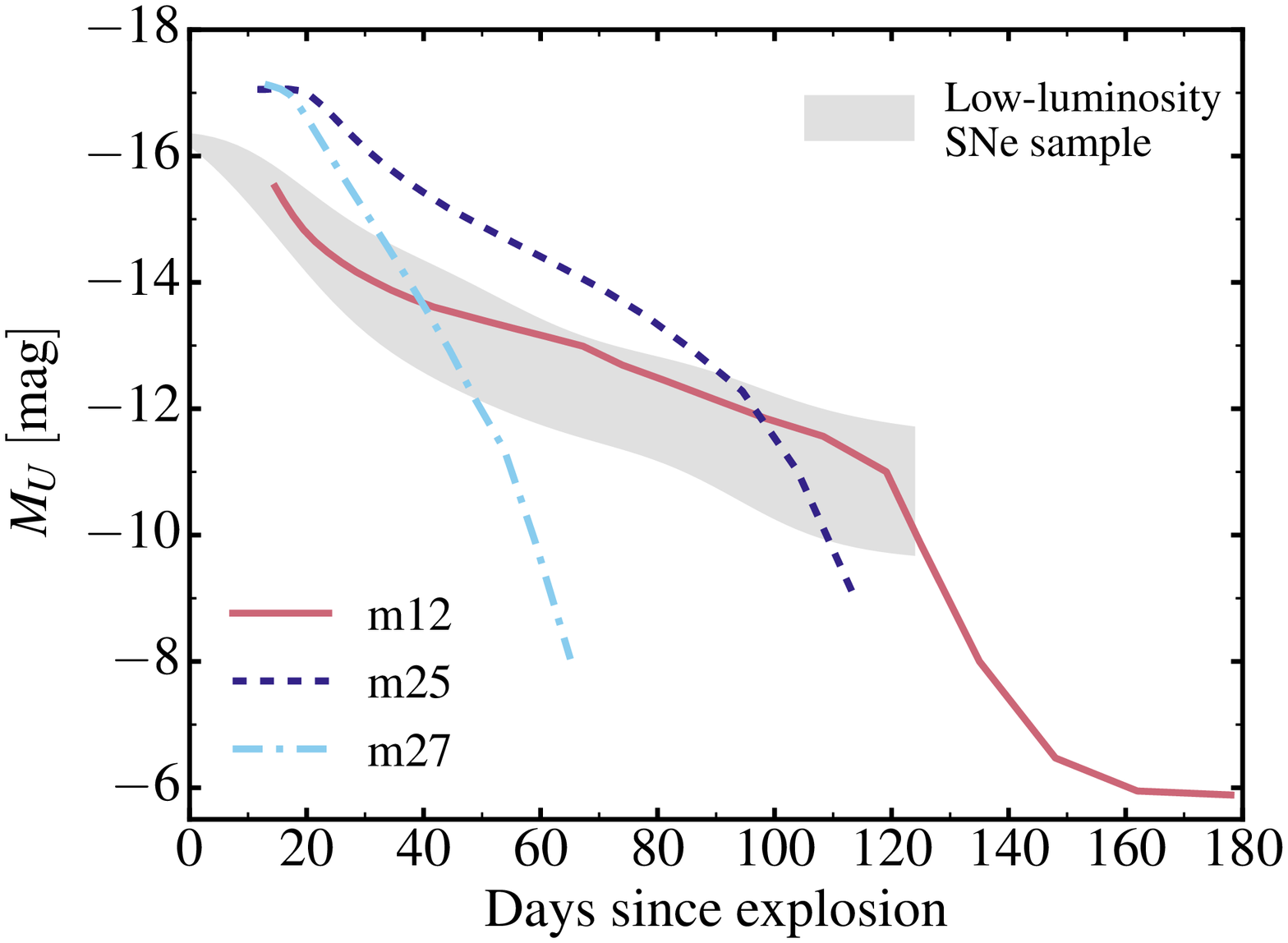}
\includegraphics[width=0.48\textwidth]{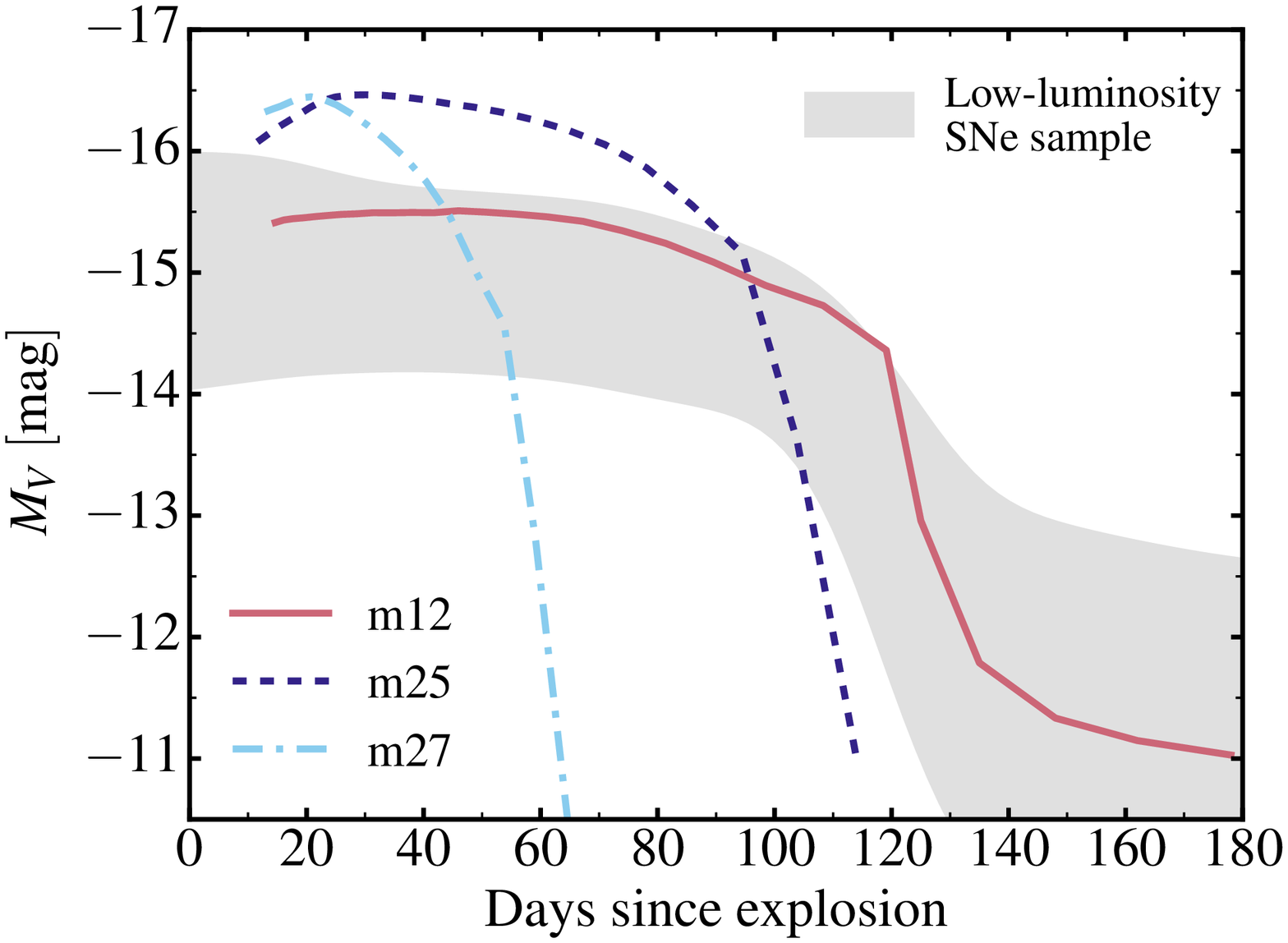}
\includegraphics[width=0.48\textwidth]{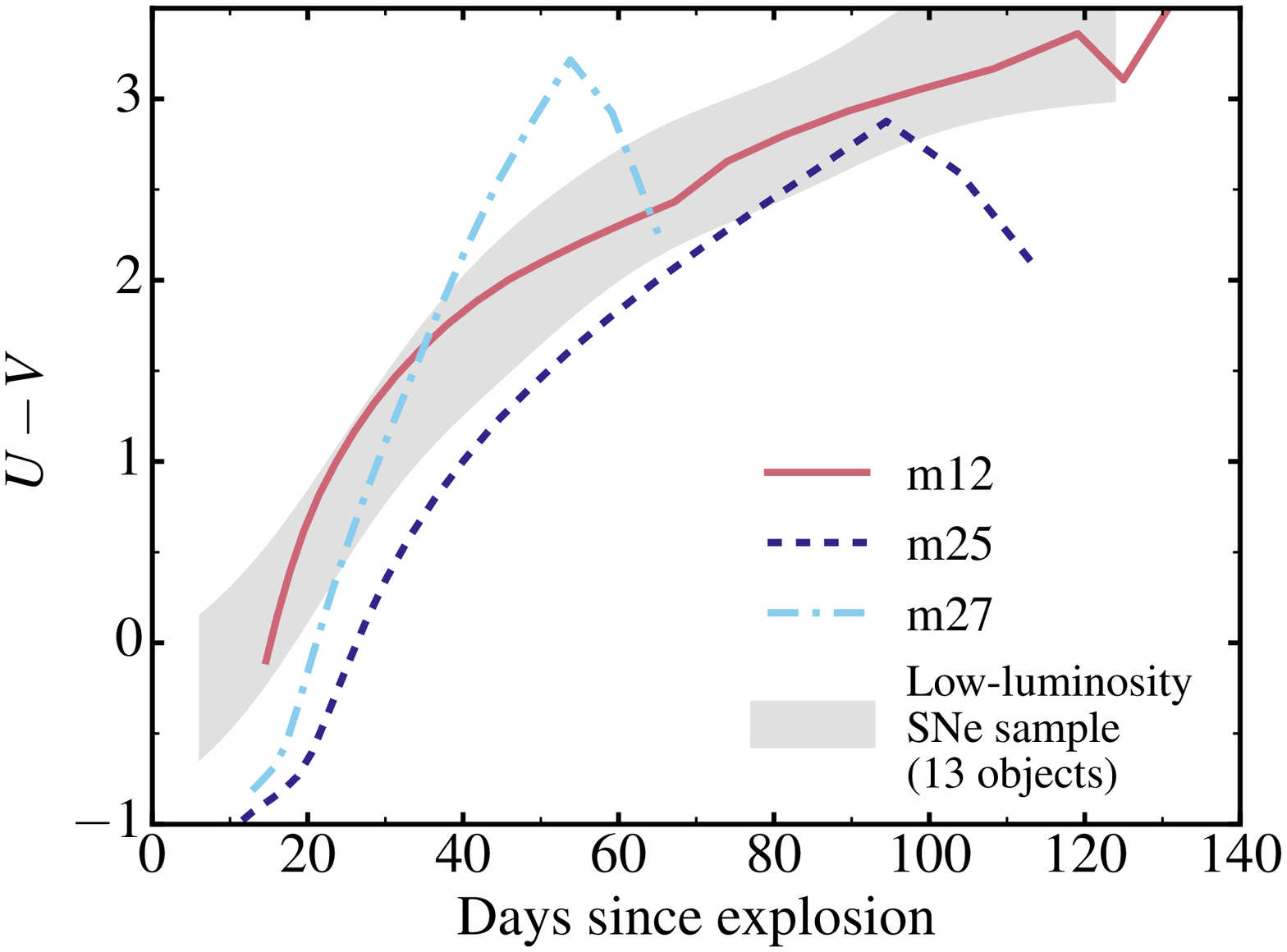}
\includegraphics[width=0.48\textwidth]{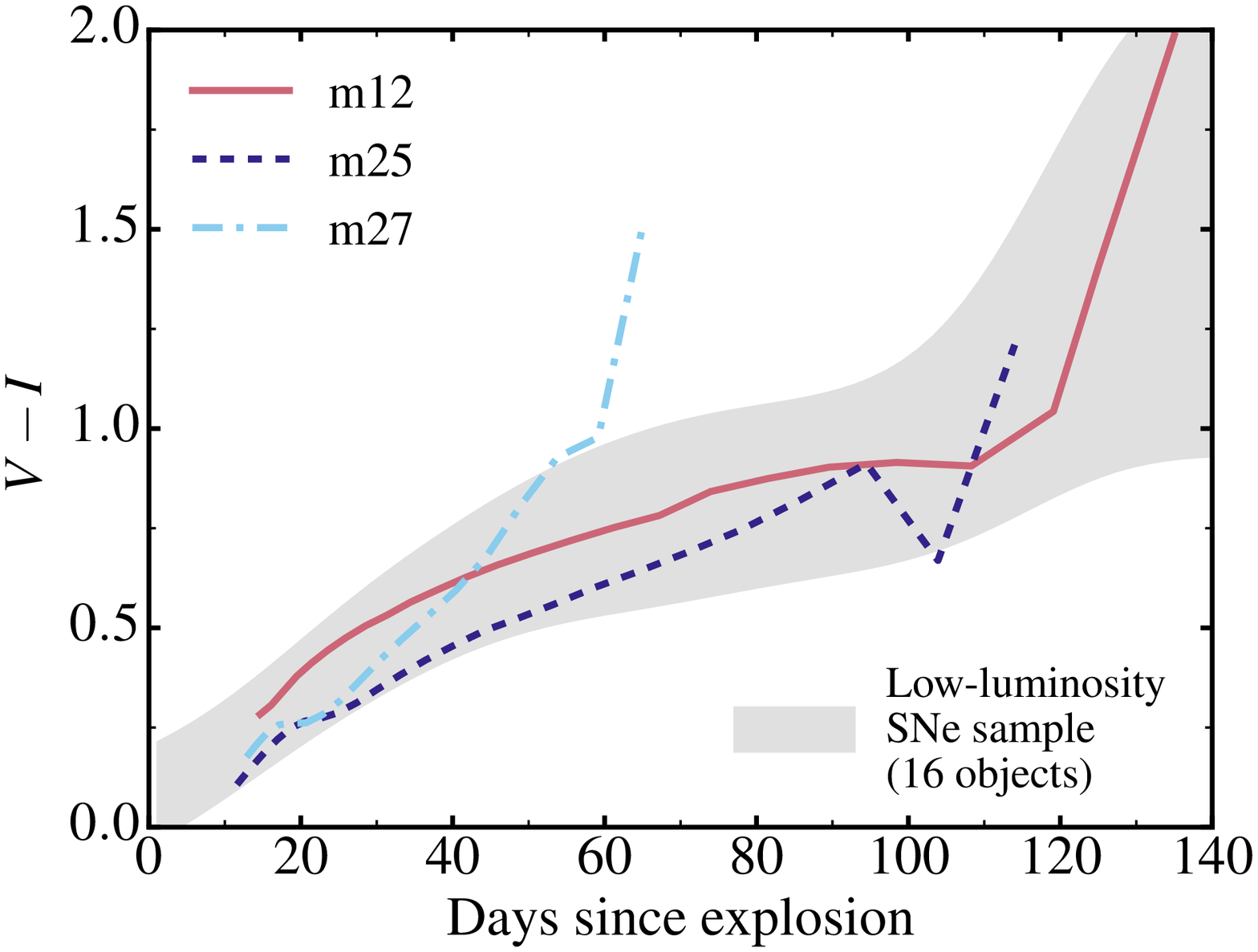}
\caption{Evolution of the \cmfgen\ $U$ and $V$ absolute magnitudes, as well as
$U-V$ and $V-I$ colors for models m12, m25, and m27.
We shade the region where the observed low-luminosity SNe II-P reside
(see Section~\ref{analysis-obs-data}). Model m12 produces the closest match
to the data.
\label{lc-models} }
\end{figure*}

\begin{figure}
\includegraphics[width=0.48\textwidth]{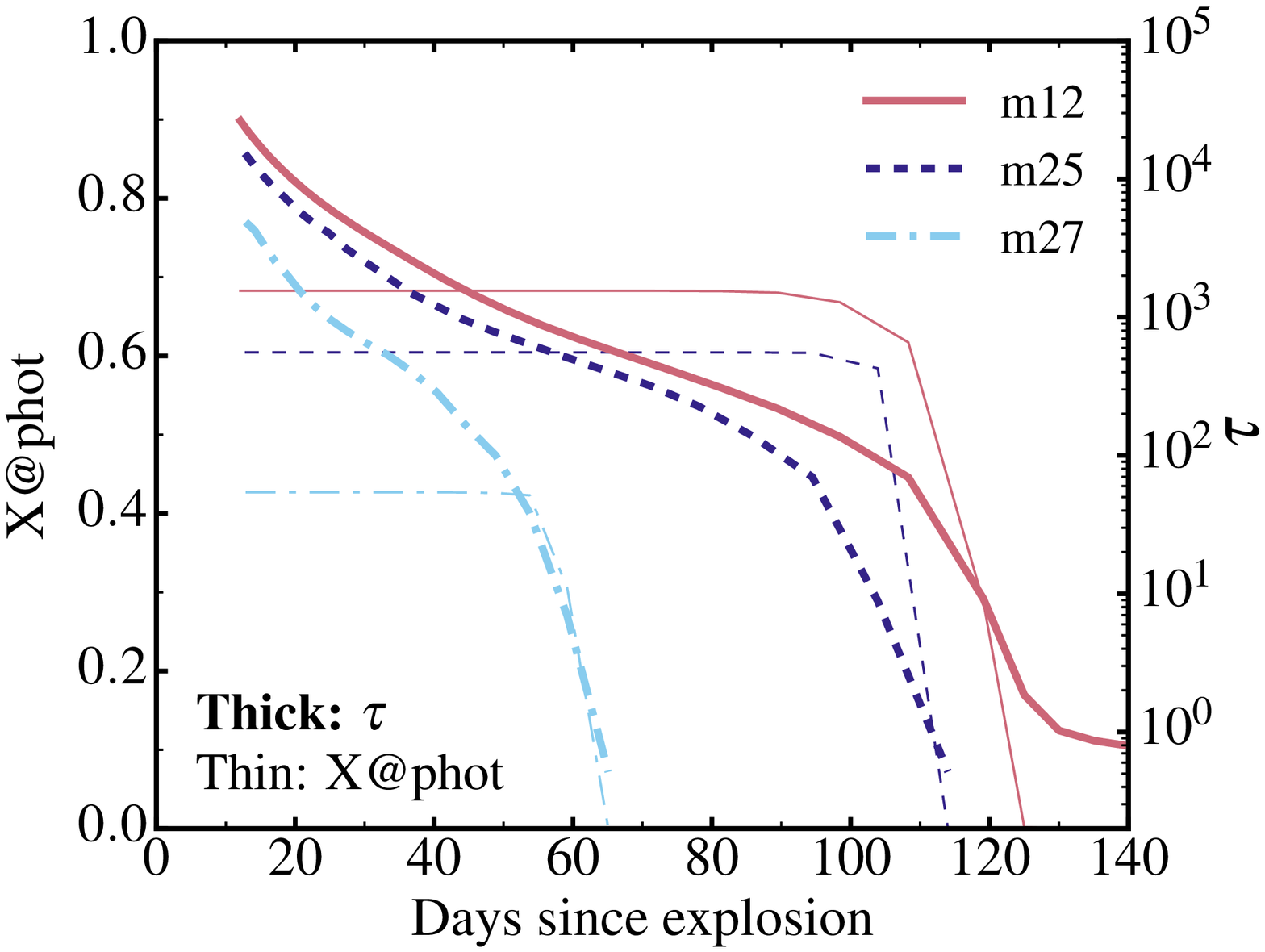}
\caption{
Evolution of the hydrogen mass fraction at the photosphere
  (thin lines) and the total ejecta electron-scattering optical depth $\tau$
  (thick lines) obtained by \cmfgen\ for models m12, m25, and m27.
\label{h-tau}}
\end{figure}

\section{Bolometric and multi-band light curves}\label{results-lc}

\subsection{Results from simulations}
Figure~\ref{fig_lbol} shows the \cmfgen\ bolometric light curves for models
m12, m25, and m27. In this order, the morphology of the bolometric light curve
goes from a plateau to a fast declining one. The faster the decline rate, the
greater the early-time luminosity, and also the earlier the transition to the
nebular phase. These properties are a consequence of the progenitor structure.
Here, the II-P/II-L morphology is largely a result of the high/low H-rich
envelope mass in the progenitor \citep{BB_2l_92}. The association of a
faster-declining light curve with a higher brightness at early times and a
shorter photospheric phase duration is a consequence of the greater $E_{\rm
kin}/M_{\rm ej}$ in model m27 compared to model m12. This correlation is observed in
the large sample of Type II SNe of \citet{Anderson2014a}.

Figure~\ref{lc-models} shows the evolution of the $U$ and $V$ band absolute
magnitudes as well as the $U-V$ and $V-I$ colors for the models m12, m25, and
m27. The morphology of the bolometric light curve discussed above is partly
reflected by these curves but not exactly because of the different color
evolution. The larger radius in models m25 and m27 (870 and 640\,\rsun) cause
bluer optical colors than in model m12 (500\,\rsun). However, the higher
$E_{\rm kin}/M_{\rm ej}$ cause a faster drop of the brightness in all optical
bands for the two higher mass models. The effect is exacerbated in model m27
because of the low H-rich envelope mass in the progenitor. This produces a
faster declining $U$-band light curve in higher mass models (they start bluer
but fade faster bolometrically). In model m25, the rise time to the brighter
$V$-band maximum is longer than in model m12 because of the bigger radius, as
obtained by \citet{DH2011} and \citet{Dessart2013}.

The $V$-band LC for model m12 shows a long plateau of $\sim$\,120\,d, which
corresponds closely to the duration of the photospheric phase (i.e., when the
ejecta electron-scattering optical depth is greater than $2/3$). For higher
mass models, the LC first rises to a maximum at $\gtrsim$\,20\,d and then
declines rapidly without showing a plateau. The duration of the photospheric
phase for models m12, m25 and m27 is 131, 112, and 64\,d. In the presence of
\ni, the  photospheric phase in models m25 and m27 would have been longer,
although physically, the strong fallback in such ejecta likely inhibits the
escape of \ni. The stark contrast between models at nebular times is thus a
reflection of the difference in \ni\ mass between m12 (0.009\,\msun) and
m25/m27 (zero).

Figure~\ref{h-tau} illustrates how the total ejecta electron-scattering
optical-depth $\tau$ and the H mass fraction at the photosphere evolve with
time. The photosphere remains in the H-rich layers until $\tau$ drops to a few
tens, after which it decreases faster. In homologous ejecta, $\tau$ evolves
as $1/t^2$ if the ionization is fixed. When the material recombines (at early
times and also at the end of the plateau), $\tau$ drops much faster (see also
\citealt{DH2011}). In model m12, $\tau$ follows a steady $1/t^2$ evolution at
nebular times because the ionization changes little (we are in a steady state
and the luminosity follows the \co\ decay rate). In models m25 and m27, the
absence of \ni\  in the ejecta makes $\tau$ (or the ionization) and the
luminosity plummet.

\begin{figure*}
\includegraphics[width=0.99\textwidth]{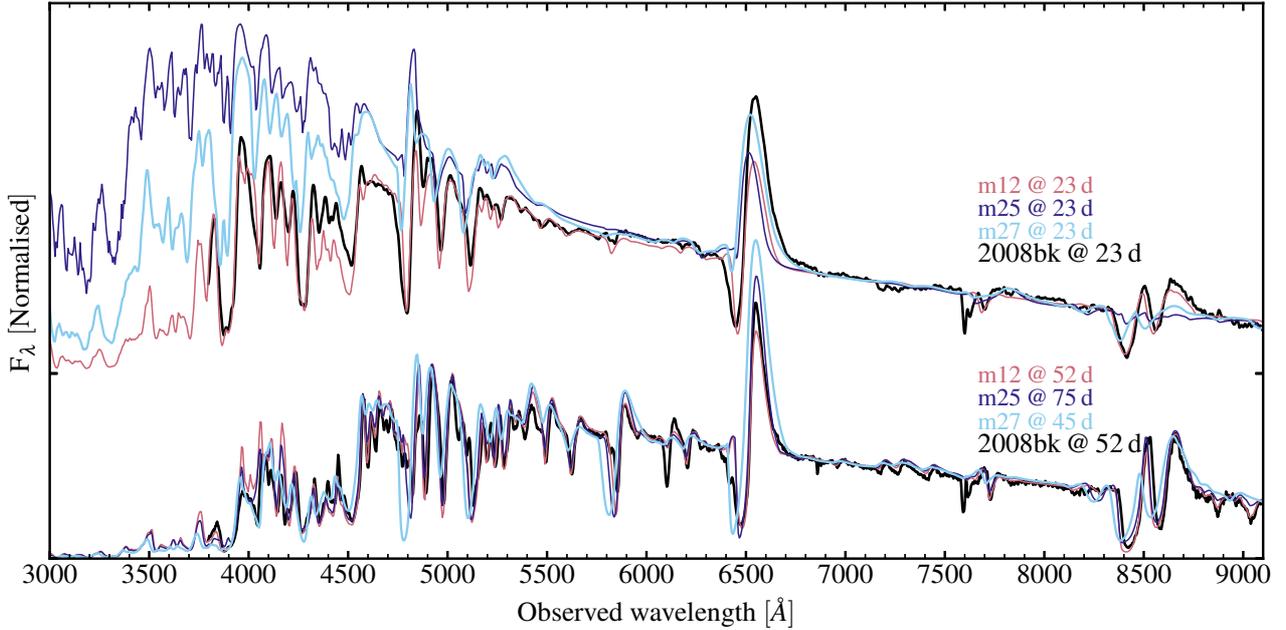}
\caption{
  Spectral comparison of models m12, m25, and m27 at 23\,d after explosion and
  when their $U-V$ color is $\sim$\,2\,mag (which corresponds to different
  post-explosion times, as indicated by the labels; see also
  Fig.~\ref{lc-models}). The ordinate ticks mark the zero flux level
  for each spectrum. We overplot the corresponding spectrum of SN\,2008bk
  (corrected for extinction and redshift). Fluxes are normalized at 7100\,\AA.
  While it is hard to distinguish the models at the recombination epoch
  (ignoring the offset in line widths, the high mass models agree with
  SN\,2008bk as well as model m12), the spectra for high mass models are
  strongly discrepant at early times. The broad absorption at 7600\,\AA\ is
  caused by atmospheric absorption.
  \label{sp-comp} }
\end{figure*}

\subsection{Comparison to observations}

In Fig.~\ref{lc-models}, the shaded area in each panel contains the scatter of
data points for the observed SNe II-P. Model m12 fits best the observed
distribution, while model m25 gives a poor match, and model m27 does not fit
the observations at all. The disagreement comes from the distinct properties of
the m25/m27 models.

While all observed low-luminosity SNe II-P exhibit a \mbox{120-d} long plateau
in their $V$-band light curve, this property is best matched by model m12
(Fig.~\ref{fig_obs_quad}). Model m25 shows roughly a $V$-band plateau (but
shorter). Model m27 has an even shorter high-brightness phase and a more
pronounced declining $V$-band light curve (model m27 shows a 0.7\,mag drop in
$V$-band magnitude between 20 and 40\,d after explosion, which is similar to
the 0.6\,mag drop that is observed in the Type II-L SN\,1979C;
\citealt{Vaucouleurs1981}). The lack of fast decliners in the observed
low-luminosity SNe II-P suggests that the progenitors have massive and dense
H-rich envelopes, which excludes a high mass progenitor like m27.

At nebular times, the \ni-deficient models m25$/$m27 do not match any
low-luminosity SN II-P, which eject at least 0.001\,\msun\ of \ni\
\citep{Pastorello2004,Spiro2014}. Mixing in \mbox{1-D} piston-driven explosions is a
parameter so it could be additionally enhanced in our high mass models to
attempt to eject a small amount of \ni. However, this would be highly
contrived. In model m12, the low production of \ni\ is a natural consequence
of the weak SN shock and the progenitor density structure (i.e., little mass
at high density). A low \ni\ mass production in a low-mass RSG is therefore
expected \citep{kitaura+06,Ugliano2012,sukhbold_16}. Although not compelling evidence, the absence of
low-luminosity SNe II-P that eject no \ni\ is suggestive that a low-mass
progenitor is more suited for these events.

At early times, the high mass models m25 and m27 also exhibit bluer colors than the observed
sample of low-luminosity SNe II-P. The effect is more striking when comparing
blue and red filters in the optical, e.g., $U-V$ rather than $V-I$, because the
peak of the spectral energy distribution is around 6000\,\AA.
The bluer colors of high mass progenitors stem primarily from their larger progenitor
radius, which arises from their larger He-core luminosity.
Although the early-time brightness can be reduced by lowering the explosion energy
(or by reducing the progenitor mass loss to produce a more massive H-rich envelope),
high-mass models will tend to be bluer and decline faster than both the lower-mass
counterparts and the observations.

\section{Spectra}\label{results-spectra}

\subsection{Results from simulations}

\begin{figure*}
\includegraphics[width=0.49\textwidth]{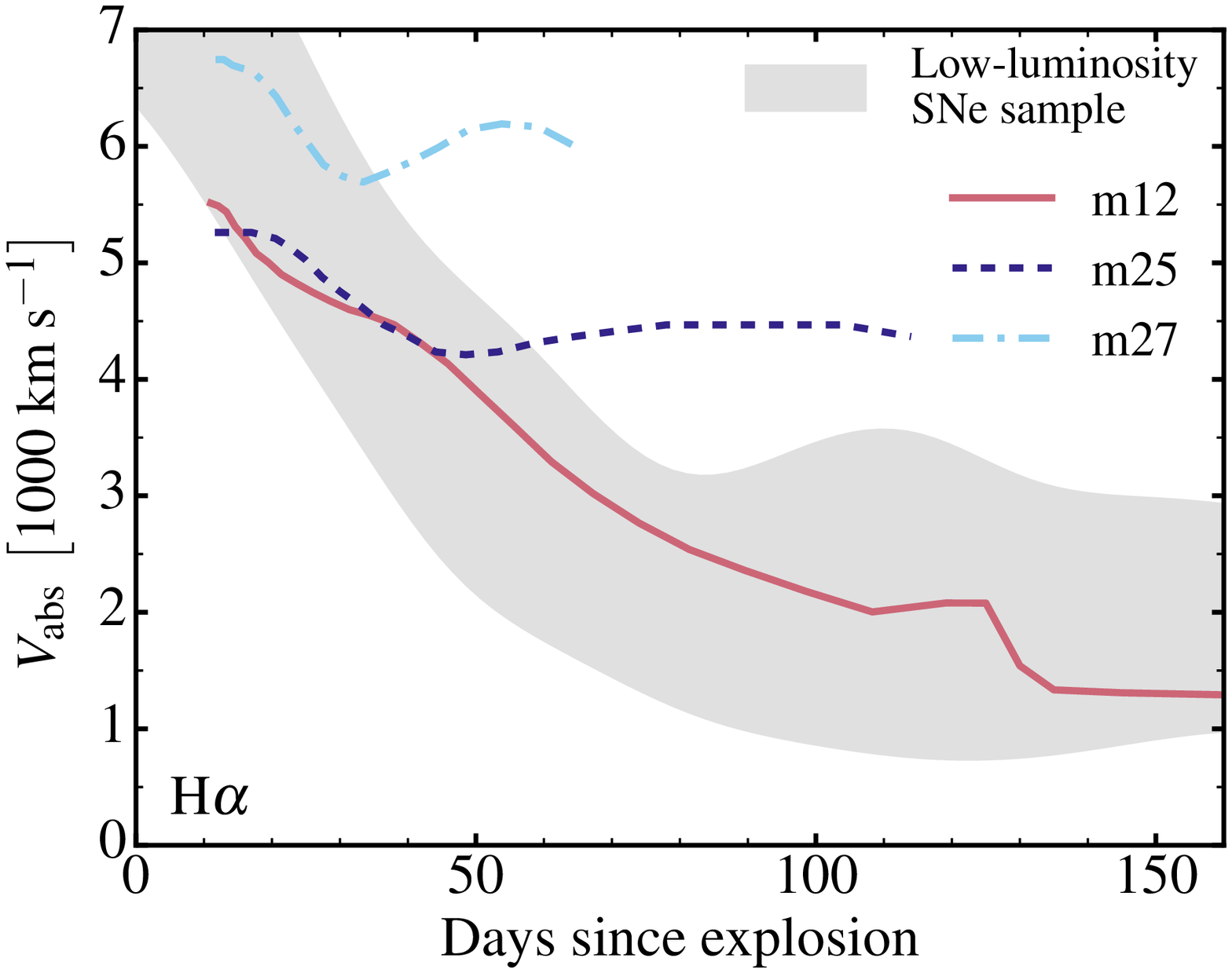}
\includegraphics[width=0.49\textwidth]{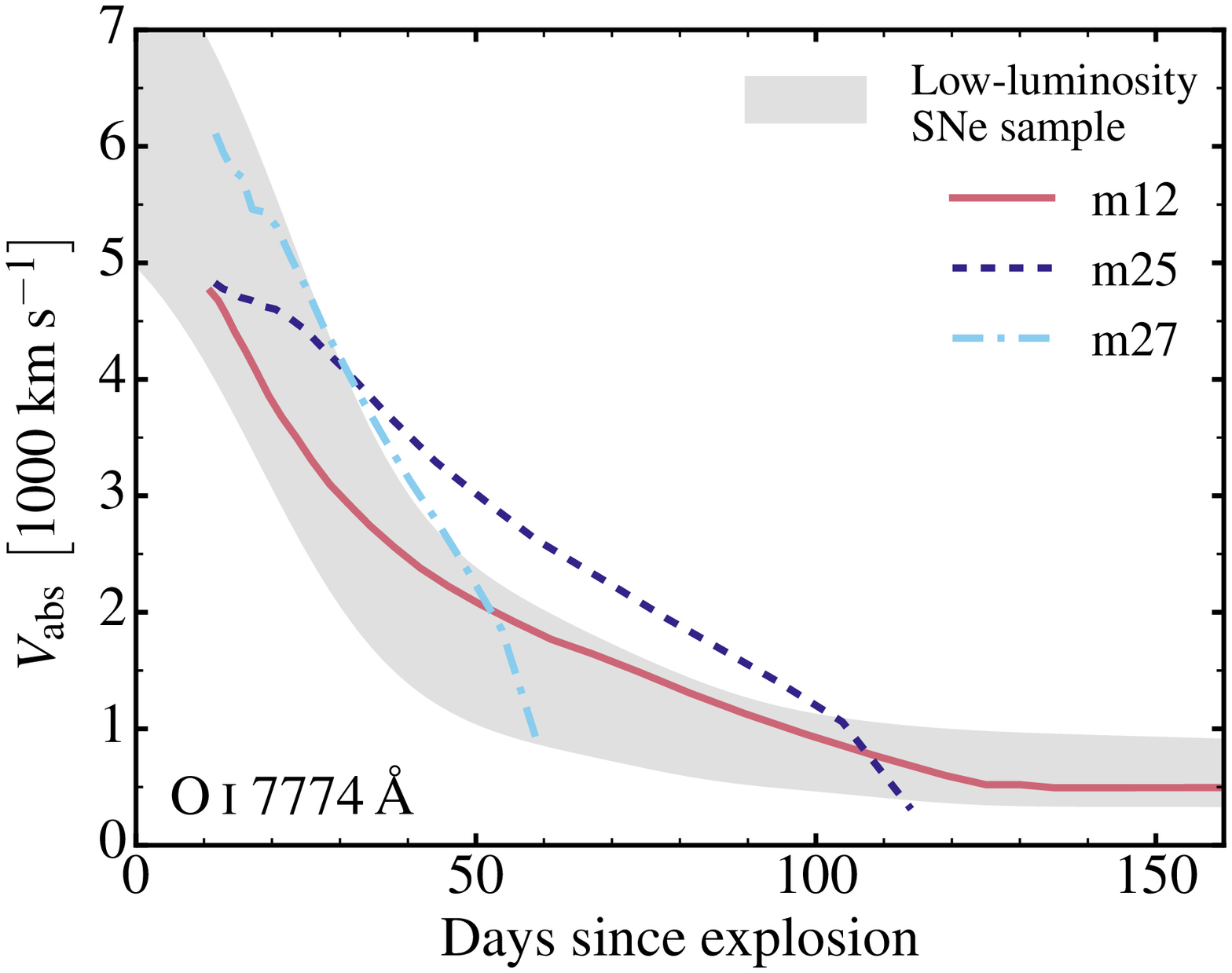}
  \caption{Velocity at absorption maximum in H$\alpha$ (left) and
    O\one\,7774\,\AA\ (right) for models m12, m25 and m27 compared
    to the observations.
    The Doppler velocity at maximum absorption in O\one\,7774\,\AA\ is a good
    tracer of the photospheric velocity (L17), while H$\alpha$ provides constraints
    on the H-rich envelope mass/kinematics.
    \label{models-vel} }
\end{figure*}

Figure~\ref{sp-comp} shows a spectral comparison for models m12, m25, and m27
at 23\,d after explosion and when $U-V \sim 2$\,mag (which corresponds
to post-explosion times of about $50-70$\,d after explosion).

At early times, the color difference discussed above is reflected in the
different spectral energy distribution. As we step from model m12, m27, to
m25, the optical spectrum is bluer, shows weaker signs of line blanketing, and
has broader lines. This directly reflects the trend in progenitor radius,
which impacts the cooling from expansion. The spectral signatures are broader
(with more line overlap) in model m27, something that arises from the similar
$E_{\rm kin}/M_{\rm ej}$ amongst models but the much lower H-rich envelope
progenitor mass in model m27.
In model m12, the H$\alpha$ and H$\beta$ line widths and strengths 
are somewhat underestimated, while the width and strength of Ca\two, 
Na\one, Sc\two, or Fe\two\ lines are well matched. 
One possible origin for the mismatch of the Balmer lines is an inadequate 
treatment of the outermost layers of the progenitor. In model X we impose
a very steep surface scale height of 0.01\,$R_{\star}$ that results in
a steep drop in the ejecta density at 5500\,\kms. A more extended 
progenitor atmosphere
would have produced a more gradual and continuous decrease in density at large
velocities, perhaps resolving this conflict. The complexity of RSG atmospheres
compromises an adequate description of these layers in our pre-explosion model.  
However, the fair agreement for all lines suggests the ejecta kinetic energy is adequate (L17).

At later times during the photospheric phase (when $U-V \sim 2$\,mag),
the spectral properties are very similar between our three models. The difference is
primarily from the width of the lines, which is somewhat greater in models m25
and m27 because of the greater $E_{\rm kin}/M_\textrm{H-env}$ relative to model
m12. The broader lines in model m27 cause greater line overlap, in particular
in the blue part of the optical where line blanketing is strongest in Type II
SNe. The effect is present in H$\alpha$, but also in metal lines like
O\one\,7774\,\AA\ (Fig.~\ref{models-vel}; the Doppler velocity at maximum
absorption in this line matches closely the photospheric velocity during the
high-brightness phase; L17). The comparison at this late epoch should be
considered with caution. Models m25 and m27 have an ejecta optical depth of
about 200, but this is about 1000 for model m12. Combined with the higher
velocities, this implies lower photospheric densities in the higher mass
models.

Overall, the synthetic spectra for our three models m12, m25, and m27, are
very similar when compared at the same $U-V$ color. When considered with respect
to the time of explosion, the offset between spectra is much greater because of the impact
of differences in progenitor radius, mean expansion rate etc., which strongly affect
the photometric (brightness and color) evolution.

The evolution of the H$\alpha$ line width can be used to distinguish the
models. Because of the different H-rich envelope to He-core mass ratio in the
models m12, m25, and m27, the velocity at the base of the H-rich envelope is
very different (despite the similar ejecta kinetic energy;
Table~\ref{mean-vel}; see also \citealt{DLW2010b}). Although mixing was
applied to all models, the big He-core in models m25 and m27 prevents much
mixing of H deep inside the He core. As a result, the H$\alpha$ line remains
broad at late times. This is in contrast to the m12 model in which most lines
progressively narrow as time progresses.

\begin{figure*}
\includegraphics[width=0.99\textwidth]{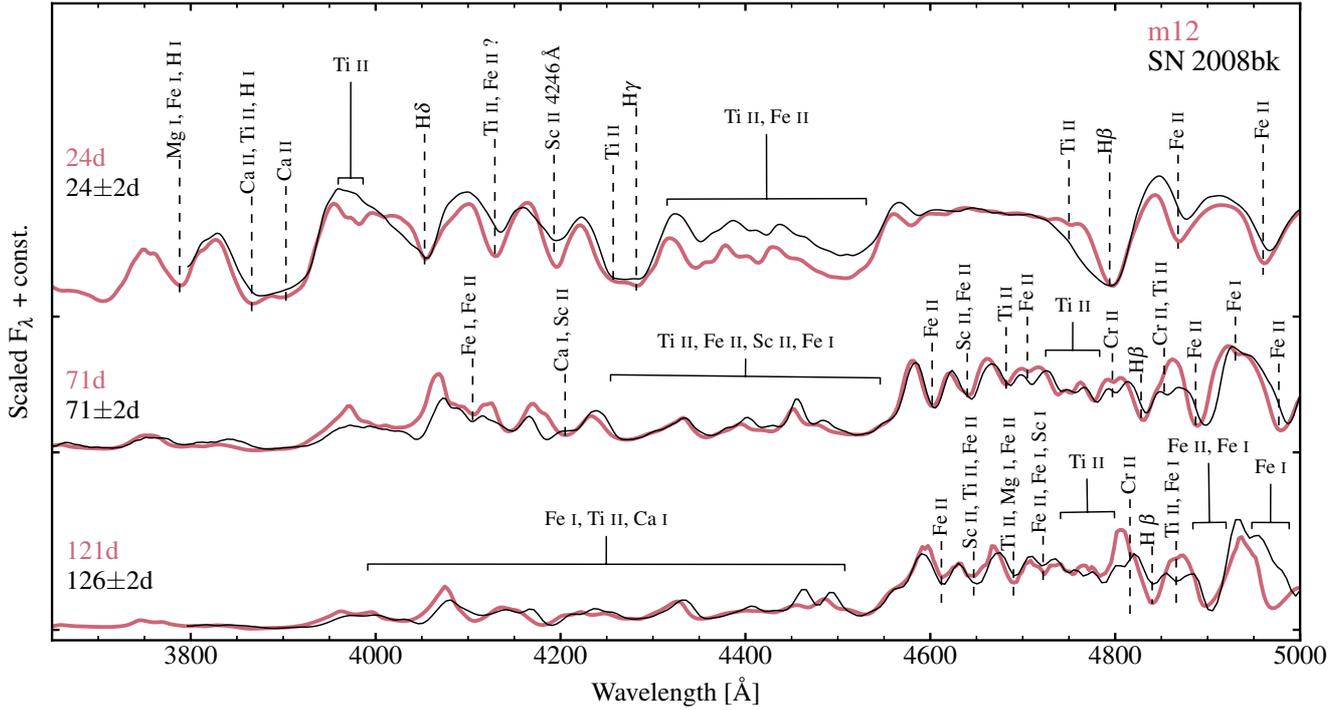}
\caption{A comparison between low-luminosity SN\,2008bk and our model m12.
  Spectra of SN\,2008bk are corrected for the redshift and the reddening.
  The ordinate ticks mark the zero flux level for each spectrum.
  Dashed lines point to the position of the absorption maximum of the
  corresponding ion for model m12. Lines are identified from the synthetic
  spectra computed without the bound-bound transitions of a given ion.
  \label{line-ids-1}}
\end{figure*}

\begin{figure*}
\includegraphics[width=0.99\textwidth]{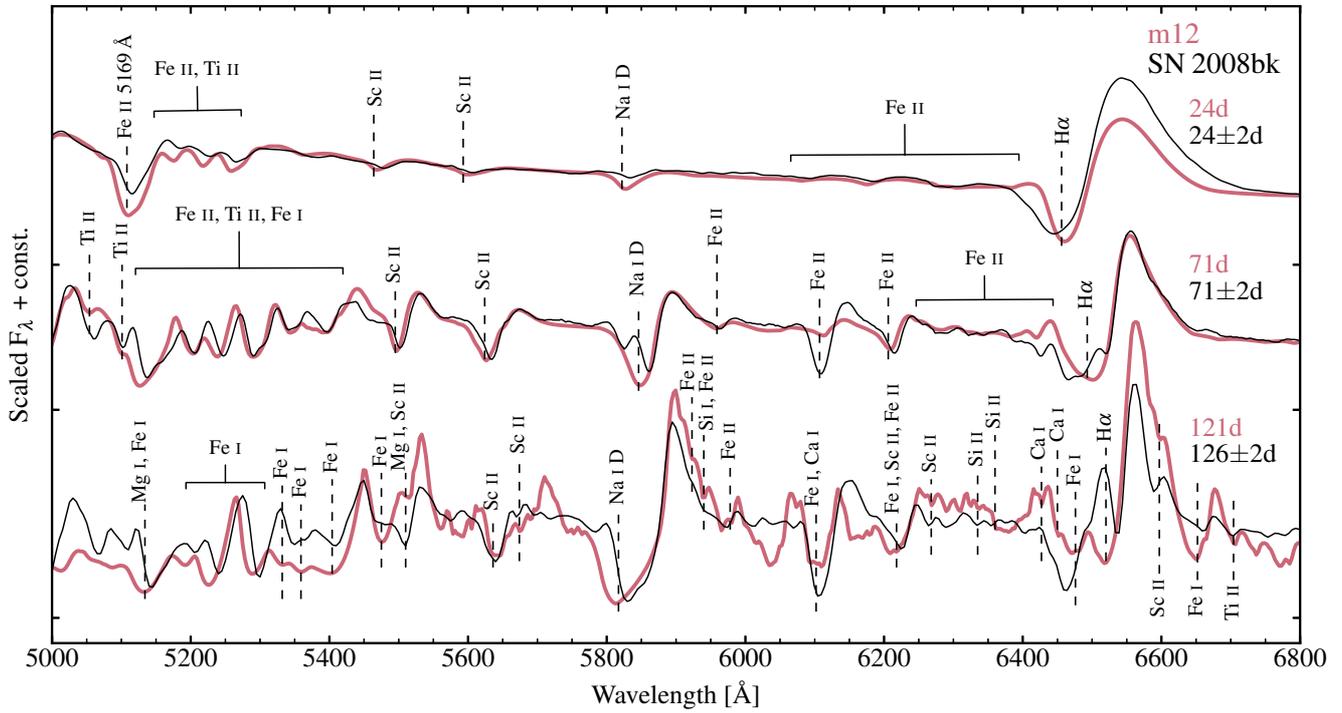}
\caption{
  Same as Fig.~\ref{line-ids-1}, but showing the 5000--6800\,\AA~region. Region
  from 6250\,\AA~to 6350\,\AA~is formed by complex and rich contributions from
  Fe\one, Fe\two, Sc\two, Si\two, O\one,
  Sc\one, and Si\one. We show in the plot only those lines that can be
  identified with certainty.
  Ba\two, which is omitted in the spectral model, is responsible for the feature at 6100\,\AA\
  and contributes to the complicated structure in the H$\alpha$ region (see L17 for details).
  \label{line-ids-2}}
\end{figure*}

\begin{figure*}
\includegraphics[width=0.99\textwidth]{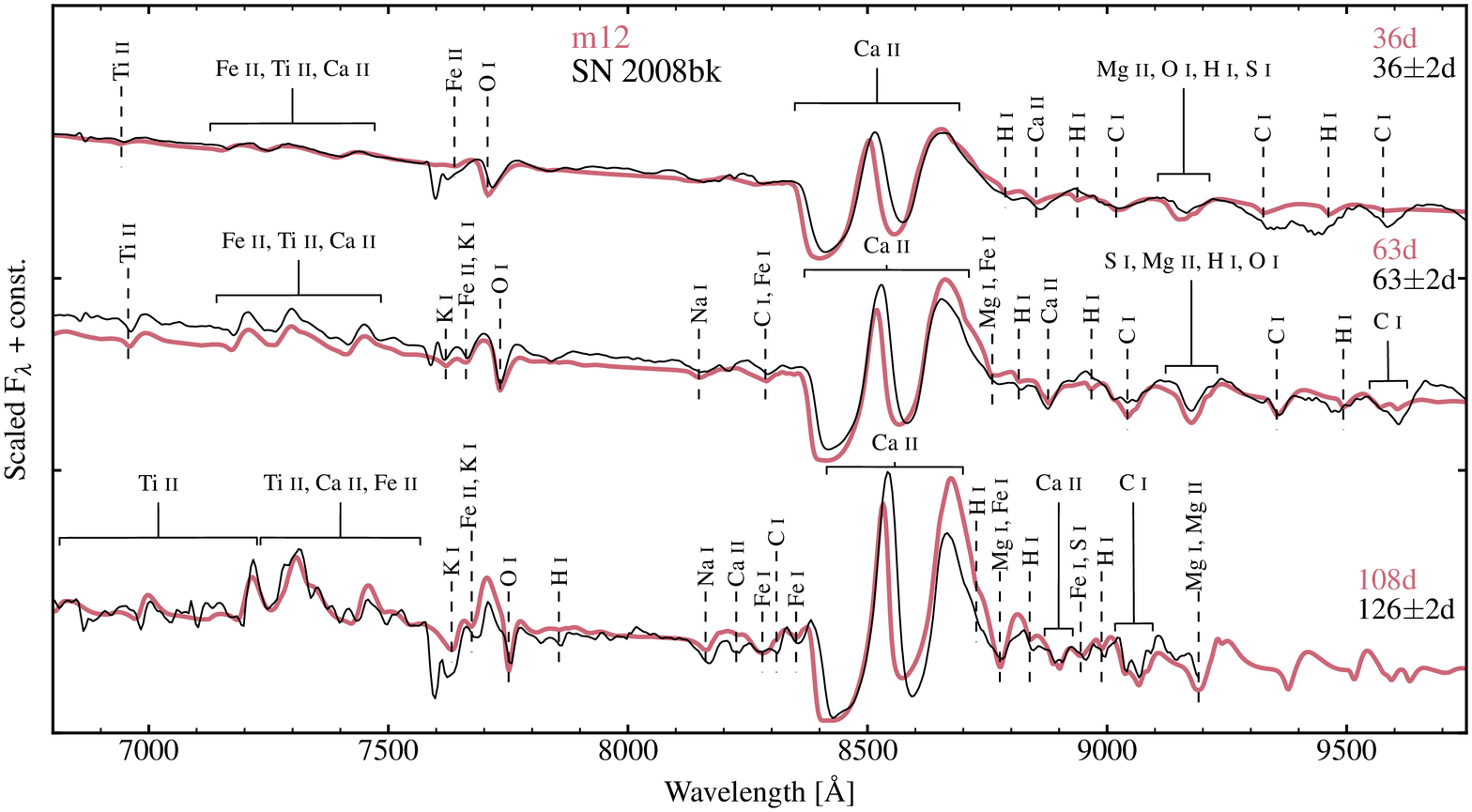}
\caption{
  Same as Fig.~\ref{line-ids-1}, but showing the 6800--9750\,\AA~region.
  The dip at $\sim$\,7600\,\AA\ is caused by atmospheric absorption.
  \label{line-ids-3}}
\end{figure*}

\subsection{Comparison to observations and spectral line identifications}\label{mod-obs-comp:sp}

In L17, we demonstrated that model m12 (named model X in L17) gave a good match to
the spectral evolution of SN\,2008bk,
which is a prototype for the sample of low-luminosity SNe II-P.
The distinct spectral evolution of models m25 and m27
relative to model m12 implies that these high mass models fail to match the spectral evolution
of SN\,2008bk, and by extension that of the whole sample of low-luminosity SNe II-P
(see Fig.~\ref{sne-montage}). The mismatch at early times is tied to the different color evolution
(which impacts the spectral index and ionization; Fig.~\ref{sp-comp}).
The evolution of the Doppler velocity at maximum absorption in H$\alpha$ also shows a
late plateau at a high value of 4000--6000\,\kms\ in models m25 and m27, while the observations
show a reduction of this Doppler velocity to very low values of $\sim$\,1000\,\kms, as obtained
in model m12 (Fig.~\ref{models-vel}).
Hence, we find that high mass models show numerous photometric and spectroscopic
discrepancies with respect to observed low-luminosity SNe II-P, while the low mass
model m12 fares better.

The lower expansion rate of low-luminosity SN II-P ejecta reduces the amount
of line overlap and facilitates line identifications. In the remainder of this section,
we discuss in more detail the spectral lines seen in our model m12 and the
high quality observations of SN\,2008bk.

Figures~\ref{line-ids-1},~\ref{line-ids-2} and~\ref{line-ids-3} show
a spectral comparison for model m12 and SN\,2008bk in three
consecutive spectral ranges spanning the optical from 3500 to 9750\,\AA,
and covering the early photospheric phase, the plateau phase, and
the beginning of the nebular phase.
Lines that we could identify are labelled in these figures. We find that all lines
observed are predicted by model m12, with just a few exceptions.
First, as reported in L17, some features in the red part of the optical in SN\,2008bk
are absent in model m12 --- this may be an instrumental artifact or an improper correction for
atmospheric absorption.
As discussed in L17, our simulations do not have Ba\two\ by default. Ba is an
s-process element and is not treated in \mesa. In L17, assuming a solar
abundance for Ba, we were able to explain a few lines blueward of H$\alpha$\ as
stemming from Ba\two\ (most notably 6141.7 and 6496.9\,\AA).
A striking feature not predicted by model m12 is the double-dip in Na\one\,D.
This double dip fits within the Na\one\,D P-Cygni trough in model m12.
It may be caused by Ba\two\,5853.7\,\AA, although our model with Ba\two\ in
L17 predicts that line to be quite weak. This feature requires further study.

\section{Comparison to other work}\label{other-work-comp}

\begin{table*}
  \caption{Summary of the inferred properties for low-luminosity SNe II-P
  ejecta and progenitors. $M_{\rm i}$ is the initial mass of the progenitor star on 
  the zero-age main sequence.
  M$_{\rm f}$ and R$_\star$ correspond the properties at the time of explosion.
  \label{llsn2p-others}}
  \begin{tabular}{
    p{0.8cm} p{1.5cm}             p{1.5cm}             p{1.5cm}       p{2.0cm}        p{2.0cm}            p{1.5cm}          p{3.3cm}  }               \hline
    SN       &$M_{\rm i}$         &$M_{\rm f}$         &$M_{\rm ej}$  &$R_\star$      &$M$(\ni)           &$E_{\rm kin}$    &Reference                \\
             &[\msun]             &[\msun]             &[\msun]       &[\rsun]        &[\msun]            &[$10^{50}$\,erg] &                         \\\hline
    1997D    &10$\pm$2            &---                 &6$\pm$1       &85$\pm$60      &0.002              &1.0$\pm$0.3      &\citealt{Chugai2000}     \\
             &---                 &26                  &24            &$\lesssim$300  &0.0025$\pm$0.0015  &4.0              &\citealt{Turatto1998}    \\
             &---                 &---                 &17            &130            &0.008              &9.0              &\citealt{Zampieri2003}   \\\hline
    1999br   &---                 &---                 &14            &108            &0.002              &6.0              &\citealt{Zampieri2003}   \\\hline
    2003Z    &14.15$\pm0.95$      &12.95$\pm$0.35      &11.3          &260            &0.005$\pm$0.003    &1.6              &\citealt{pumo_2p_17}     \\
             &15.9$\pm1.5$        &---                 &14$\pm$1.2    &229$\pm$39     &0.0063$\pm$0.0006  &2.45$\pm$0.18    &\citealt{Utrobin2007b}   \\\hline
    2005cs   &---                 &---                 &10.5$\pm$2.5  &100            &0.003              &3.0              &\citealt{Pastorello2009} \\
             &18.2$\pm$1          &17.3$\pm$1          &---           &600$\pm$140    &0.0082$\pm$0.0016  &4.1$\pm$0.3      &\citealt{Utrobin2008}    \\\hline
    2008bk   &12                  &9.88                &8.29          &502            &0.0086             &2.5              & L17                     \\
             &12.15$\pm0.75$      &11.65$\pm$0.35      &10.0          &503            &0.007$\pm$0.001    &1.8              &\citealt{pumo_2p_17}     \\\hline
    2008in   &20                  &---                 &16$\pm$4      &---            &0.025$\pm$0.01     &8.6$\pm$2.5      &\citealt{Gurugubelli2011}\\
             &$\leq20$            &---                 &16.7          &126            &0.015              &5.4              &\citealt{Roy2011}        \\
             &---                 &15.5$\pm$2.2        &13.6$\pm$1.9  &570$\pm$100    &0.015$\pm$0.005    &5.05$\pm$3.4     &\citealt{Utrobin2013}    \\\hline
    2009N    &---                 &13.25$\pm$0.25      &11.5          &287            &0.02$\pm$0.004     &4.8              &\citealt{Takats2014}     \\\hline
    2009md   &12.15$\pm0.75$      &11.65$\pm$0.35      &10.0          &287            &0.004$\pm$0.001    &1.7              &\citealt{pumo_2p_17}     \\
             &---                 &8.5$^{+6.5}_{-1.5}$ &---           &500            &0.0054$\pm$0.003   &---              &\citealt{Fraser2011}     \\\hline
  \end{tabular}
\end{table*}

Our results for the $V$-band magnitude and the photospheric velocity at 50\,d
after explosion in model m12 (--15.61\,mag, 2401\,\kms)
are in rough agreement with those of \citet{KW2009} for their model
M15\_E0.3\_Z1 of comparable ejecta energy and mass (--15.96\,mag and 3125\,\kms).
The plateau duration of 120\,d in model m12 is, however, unmatched by any of the
low energy models in \citet{KW2009}, which are all longer than 150\,d. This likely
arises from the large \ni\ mass in their simulations, which exceeds 0.1\,\msun\ in all cases.
The \ni\ mass distribution of their model set is 1-2 orders of magnitude larger than
the value inferred for low-luminosity SNe II-P.

Models m25 and m27 do not eject any \ni\ because of the strong fallback in those
progenitors. In contrast, the models of \citet{KW2009} eject a significant amount of \ni\ even
in the low energy explosions of high mass progenitors (e.g., 0.34\,\msun\ in their model
M25\_E0.6\_Z1). The reason for this difference is unclear. In the simulations of \citet{DLW2010b},
most of the CO core falls back if the progenitor mass is $\gtrsim$\,20\,\msun\ and the ejecta
kinetic energy at infinity is 3$\times$10$^{50}$\,erg. The smaller the piston power, the greater
is the fallback. So, the large \ni\ mass obtained in \citet{KW2009}, which is well above the inferred
value for SNe II-P, may result partly from overestimating its power.

In Table~\ref{llsn2p-others} we present the ejecta/progenitor properties inferred
from radiation-hydrodynamics modeling and/or pre-explosion photometry
of low-luminosity SNe II-P.
As discussed in the introduction, there is a large scatter in progenitor masses (but
also surface radii etc.). In this work, we have studied the whole sample of low-luminosity
SNe II-P and emphasized what a uniform set they form in terms of $V$-band LC, color
evolution, spectral evolution, or expansion rates (Section~\ref{analysis-obs-data}).
It is hard to understand how a wide range of ejecta/progenitor properties
can arise from such a uniform set of events.

Our studies favor low-mass massive stars as progenitors of SNe II-P, which is
in agreement with \citet{Pastorello2004}, \citet{Spiro2014}, \citet{pumo_2p_17},
or \citet{Fraser2011}.
In some studies, the progenitor radius is claimed to be as low as  85--130\,\rsun\
(\citealt{Chugai2000, Zampieri2003, Pastorello2009, Roy2011}), which is more
typical of BSG progenitors. Low-luminosity SNe II-P do not have a Type II-pec evolution
like SN\,1987A, and such small radii are also in strong disagreement with the constraints
from pre-explosion images. Stellar evolution also predicts that the majority of low/moderate
mass massive stars die as RSG stars, not BSG stars.

Our results are in conflict with the results of \citet{Turatto1998}, who propose
a 24\,\msun\ ejecta for SN\,1997D.

\section{Conclusions}\label{conclusions}

We have studied the properties of observed low-luminosity
SNe II-P and confronted them to the radiation properties obtained
numerically from the explosion of low- and high-mass RSG stars (12,
25 and 27\,\msun\ on the main sequence).

Observations of low-luminosity SNe II-P reveal a very uniform class of
objects, both photometrically and spectroscopically. All events show a plateau
LC in the $V$ band during the photospheric phase -- there are no fast decliners (II-L like) in this set.
The plateau duration is tightly centered around 110\,$\pm$\,10\,d.
Their color evolution is also similar, showing a progressive
and monotonic reddening during the photospheric phase. A larger scatter in
color appears at nebular times, driven from differences in \ni\ mass and
perhaps from chemical mixing in the He core (L17). All low-luminosity SN II-P
ejecta contain some \ni, with a minimum inferred mass of 0.001\,\msun.
Spectroscopically, low-luminosity SNe II-P systematically exhibit narrower
lines than standard-luminosity SNe II-P, which implies lower ejecta expansion
rate. It thus appears that low-luminosity SNe II-P are low energy explosions
of RSG stars.

Using stellar evolution and explosion models for stars of initial mass of 12, 25,
and 27\,\msun, we study the radiation properties of SNe arising from the explosion
of low- and high-mass RSG stars. We find systematic differences in SN properties 
between these two mass domains, which arise from their distinct pre-SN structure.

RSG stars of greater initial mass produce heavier He cores and greater
surface luminosities, giving rise to a greater mass loss.
Consequently, the RSG radius increases with the main-sequence mass
while the ratio of the H-rich envelope mass to the He-core mass decreases.
For large enough mass loss, the envelope may shrink, as in our model m27.
For models m12, m25, and m27, the surface radius is 502, 872, and 643\,\rsun.
As reported in \citet{DH2011,Dessart2013}, we find that the larger the progenitor
radius, the bluer the SN prior to the recombination phase.
Only the explosion of more compact, i.e., lower mass RSG stars, matches
the color evolution of low-luminosity SNe II-P.

Because the ratio of the H-rich envelope mass to the He-core mass decreases
with increasing main-sequence mass, low- and high-mass RSG stars have
a very different chemical stratification in mass space. This stratification is visible in velocity
space after explosion, with the H-rich ejecta layers being confined to higher velocity
regions in SNe II models from higher mass RSG stars.
The smaller H-rich envelope mass in higher mass RSG stars tends to produce a shorter
plateau (models m25 and m27).
As the H-rich envelope drops to just a few solar masses, the $V$-band light curve
shows a faster decline rate, in contradiction to observations of low-luminosity SNe II-P.
Furthermore, only in model m12 do H$\alpha$ and Fe\two\,5169\,\AA\ follow their observed
counterparts. In model m25 and m27, H$\alpha$ remains much too broad at late times,
reflecting the large velocity of the H-rich layers (or the large velocity of the former
He-core material).

Third, some difficulties with high-mass progenitors arise concerning the
amount of~\ni\ ejected in the explosions. This parameter is very well
constrained from the observed LCs at nebular times (in the sense that
it does not require radiative transfer modelling), and found to
be at least 0.001\,\msun. SN\,1999eu may have ejected even less \iso{56}Ni
but the nebular phase photometric data is too sparse to say confidently.
Admittedly, for very low \iso{56}Ni yields, it can become a challenge to
extract the SN brightness from the image photometry, especially for SNe lying
in relatively dense star clusters.

In our models m25 and m27, no \ni\ is ejected due to the highly bound He-core
and the strength of the reverse shock. These models experience strong fallback,
the entire CO core falling into the compact remnant and forming a $\sim$\,6\,\msun\
black hole.
The ejecta kinetic energy of $4.2\times10^{50}$\,erg in models m25 and m27,
small enough to prevent \ni\ ejection in these ejecta, is likely overestimated
as the line profiles are broader than observed during the photospheric phase.
Reducing the discrepancy in line widths at early time would require reducing the
explosion energy, which would enhance the amount of fallback, this time perhaps of the
entire He core.
 In this context, observing narrow [O\one] or [Ca\two] lines at nebular times
in a low-luminosity SN II-P is unambiguous evidence that some \iso{56}Ni is ejected.
If the power at nebular times comes instead from interaction with the progenitor wind,
only a broad H$\alpha$ line should be seen. This may help refine the interpretation
for the origin of the faint brightness at nebular times.

According to the results of our modeling and the confrontation to observed
LCs and multi-epoch spectra,  we conclude that low-mass RSG stars are
the preferred progenitor population for the observed  low-luminosity SNe II-P.
It is however unclear whether all low mass RSG stars produce
low energy explosions.
For a standard initial mass function with exponent 2.3, 42\% of massive stars
are born in the range $8-12$\,\msun, and 15\% in the range $8-9$\,\msun. In contrast,
only 5\% of all Type II SNe are low energy explosions. So, either the mass range
for these low-energy SNe II-P is very narrow (e.g., narrower than 8 to 9\,\msun),
or $8-12$\,\msun\
exhibit some diversity in explosion energy, or we are missing numerous low-energy
Type II-SNe because of an observational bias. A combination of all three might hold in Nature.
From the point of view of the explosion mechanism, low-energy explosions seem
to naturally occur in massive star progenitors characterized by a steeply declining
density profile above the degenerate core, which is a generic feature of the lowest
mass massive stars \citep{kitaura+06}. The low-energy
SNe II-P may then arise from the collapse of the ONeMg core leading to electron-capture SNe
\citep{poelarends+08}.

\section{Acknowledgements}

Sergey Lisakov is supported by the Erasmus Mundus Joint Doctorate Program 
by Grants Number 2013-1471 from the agency EACEA of the European Commission.
DJH acknowledges support from STScI theory grant HST-AR-12640.01
and NASA theory grant NNX14AB41G. 
Some of the data used in this work were downloaded from the Open Supernova Catalog
(\href{https://sne.space}{https://sne.space}, \citealt{Guillochon2017}).
This work used computing resources of the mesocentre SIGAMM, hosted by the 
Observatoire de la C\^ote d'Azur, Nice, France. 


\appendix
\section{Observational data}\label{appendix:obs-data}

Our sample consists of 17 SNe (Table~\ref{sn-data}):
1997D,
1999br,
1999gn,
2001dc,
2004eg,
2005cs,
2006ov,
2008bk,
2008in,
2009N,
2009md,
2010id,
2013am.
Some of the objects from our sample, such as 2005cs and 2008bk, were followed
spectroscopically and photometrically with a high cadence. Others have been
observed just a few times, but all these SNe are confirmed as low-luminosity Type
II-P SNe.

In Table~\ref{other-sn-data}, we summarize the main information for the SNe
II used for comparison.
The quoted extinction A$_V$ corresponds to the Galactic extinction \citep{Schlafly2011}.
However, in some cases, the A$_V$ value includes the host galaxy extinction. Such cases
are marked in the Table~\ref{other-sn-data} and discussed explicitly in below.
We adopt a visual extinction to reddening ratio \mbox{R$_V = {\rm A}_V/{\rm E}(B-V) = 3.1$}.
All objects are associated with large spiral galaxies, and are generally located within their arms.
Some relevant information not included in the Table~\ref{other-sn-data} is presented
below for each object.

\begin{table*}
  \caption{Observational data for our sample of low-luminosity Type II-P SNe.
    $V_{\rm rec}$ is the recessional velocity. See additional information in Section~\ref{appendix:obs-data}.
    \label{sn-data}}

  \begin{tabular}{
    p{1.0cm}  p{1.4cm}      p{3.0cm}                        p{2.0cm}            p{1.3cm}      p{1.6cm}          p{1.5cm}        p{1.7cm}        }  \hline
    SN        &Host galaxy  & Galaxy Type &Explosion date     &A$_V^a$      &$\mu$            &$V_{\rm rec}$  &References$^b$ \\ \hline
              &             &                               &MJD                &[mag]        &[mag]            &[\kms]         &               \\ \hline
    1994N     &UGC\;5695    &Sb                             &49451.0$\pm$10     &0.103        &33.09$\pm$0.31   &2940           &P04            \\
    1997D     &NGC\;1536    &SBc, interacting galaxies      &50361.0$\pm$15     &0.058        &31.29            &1461           &P04, S14       \\
    1999br    &NGC\;4900    &SBc                            &51278.0$\pm$3      &0.065        &31.90$^{**}$     & 968           &P04, S14       \\
    1999eu    &NGC\;1097    &SBb, AGN                       &51394.0$\pm$15     &0.073        &31.08            &1273           &P04, S14       \\
    1999gn    &M\,61        &SBbc, Seyfert 2 galaxy         &51520.0$\pm$10     &0.061        &30.50$\pm$0.20   &1616           &S14            \\
    2001dc    &NGC\;5777    &Sb                             &52047.0$\pm$5      &1.250$^{*}$  &33.19$\pm$0.43   &2140           &P04, S14, So14 \\
    2002gd    &NGC\;7537    &Sbc, interacting galaxies      &52552.0$\pm$2      &0.184        &32.87$\pm$0.35   &2678           &S14, W10       \\
    2003Z     &NGC\;2742    &Sc                             &52665.0$\pm$4      &0.106        &31.70$\pm$0.60   &1280           &S14, H12       \\
    2004eg    &UGC\;3053    &Sc                             &53170.0$\pm$30     &1.237$^{*}$  &32.64$\pm$0.38   & 2414           &S14, C11       \\
    2005cs    &M\,51        &SABb, Seyfert 2 galaxy         &53547.6$\pm$0.5    &0.124$^{*}$  &29.75$\pm$0.16   & 466           &D08            \\
    2006ov    &M\,61        &SBbc,  Seyfert 2 galaxy        &53974.0$\pm$6      &0.061        &30.50$\pm$0.20   &1616           &S14            \\
    2008bk    &NGC\;7793    &Scd                            & 54546.0$\pm$2      &0.062$^{*}$  &27.68$\pm$0.05   & 283           &L17, P10, P    \\
    2008in    &M\,61        &SBbc, Seyfert 2 galaxy         &54825.0$\pm$1      &0.305$^{*}$  &30.50$\pm$0.20   &1616           &R11            \\
    2009N     &NGC\;4487    &SBc                            &54848.1$\pm$1.2    &0.403$^{*}$  &31.67$\pm$0.11   &1050           &T14            \\
    2009md    &NGC\;3389    &Sc           &               55170.0$\pm$4    &0.380$^{*}$  &31.64$\pm$0.21   &1298           &F11, H12       \\
    2010id    &NGC\;7483    &SBa                            &55452.0$\pm$2      &0.167        &33.15$\pm$0.45   &4940           &G11, T07       \\
    2013am    &M\,65        &SBa, AGN                       &56372.0$\pm$1      &1.767$^{*}$  &30.54$\pm$0.40   & 807  &Z14, N11       \\ \hline
    1999em    &NGC\;1637    &SAB(rs)c                       &51474.3$\pm$2      &0.31$^{*}$   &30.30$\pm$0.17   & 800           &DH06, L03      \\ \hline
  \end{tabular}

  \begin{minipage}{\textwidth}

    $^a$ In most cases visual extinction A$_V$ corresponds to the Galactic
      extinction \citep{Schlafly2011}, but for some SNe A$_V$ also includes
      additional extinction (see Appendix~\ref{appendix:obs-data} for details).
      In this case A$_V$ value is followed by an asterisk. In some cases A$_V$
      is calculated from A$_B$ or $E(B-V)$, provided in the corresponding
      papers.

    $^b$
      P04:  \cite{Pastorello2004};
      S14:  \cite{Spiro2014};
      L03:  \cite{Leonard2003}; 
      DH06: \cite{DH2006};      
      D08:  \cite{Dessart2008}; 
      L17: in L17 we adopted an explosion date of 54546.0\;MJD
        for SN\;2008bk basing on the explosion date estimation of
        54548.0$\pm$2\;MJD from \cite{Pignata2013} and spectral evolution of
        model m12;
      P10: \cite{Pietrzynski2010}; 
      P:  Pignata, private communication; 
      R11: \cite{Roy2011};      
      T14: \cite{Takats2014};   
      F11: \cite{Fraser2011};   
      G11: \cite{Gal-Yam2011};  
      Z14: \cite{Zhang2014};    
      So14: \cite{Sorce2014};    
      W10: \cite{Wei2010};      
      H12: \cite{Hakobyan2012}; 
      C11: \cite{Cappellari2011}; 
      T07: \cite{Theureau2007}; 
      N11: \cite{Nasonova2011}. 

    ** The distance to SN\;1999br in \cite{Pastorello2004} is 17.3 Mpc (distance
      modulus $\mu=31.19$). \citet{Pignata2013} supposes that the distance for
      the 1999br may be underestimated, basing on the similarities between
      SN\;1999br and SN\;2008bk. We use the mean result from the NED catalogue,
      derived from 7 distance estimations to the host galaxy NGC\;4900.
  \end{minipage}
\end{table*}

\begin{table*}
  \caption{Observational data for Type II SNe, not in our low-luminosity sample, but used in the paper.
    \label{other-sn-data}}

  \begin{tabular}{
    p{1.0cm}  p{1.7cm}       p{1.7cm}        p{2.0cm}          p{1.3cm}      c                  r               p{3.7cm}        }  \hline
    SN        &Host galaxy   & Galaxy type
                                             &Explosion date   &A$_V^a$      &$\mu$             &$V_{\rm rec}$  &References$^b$ \\ \hline
              &              &               &MJD              &[mag]        &[mag]             &[\kms]         &               \\ \hline
    1969L     &NGC\;1058     &SA(rs)c        &40549$\pm$5      &0.163        &30.00$\pm$0.22    & 518           & C71, A76, L12, T88  \\
    1992ba    &NGC\;2082     &SB(r)b         &48888.5$\pm$8    &0.156        &31.34$\pm$0.53    &1246           & H01, A14, O10, F96  \\
    1999bg    &IC\;758       &SB(rs)cd?      &51251$\pm$14     &0.052        &32.41$\pm$0.18    &1290           & F14, P09, V91       \\
    1999gi    &NGC\;3184     &SAB(rs)cd      &51518$\pm$4      &0.651*       &30.34$\pm$0.14    & 592           & F14, J09, S92       \\
    2001X     &NGC\;5921     &SB(r)bc        &51963$\pm$5      &0.106        &31.85$\pm$0.22    &1470           & F14, R14            \\
    2001hg    &NGC\;4162     &(R)SA(rs)bc    &52260$\pm$20     &0.1          &33.07$\pm$0.50    &2580           & S09, F14            \\
    2002ca    &UGC\;8521     &(R)SB(r)ab pec &52353$\pm$15     &0.063        &33.03$\pm$0.45    &3264           & F14, T07            \\
    2003T     &UGC\;4864     &SA(r)ab        &52645$^c$        &0.084        &35.21$\pm$0.42    &8373           & G16, A14, O10       \\
    2003bn    &PGC\;831618   &?              &52694.5$\pm$3    &0.174        &33.67$\pm$0.42    &3831           & A14, O10            \\
    2003gd    &M\,74         &SA(s)c         &52720$\pm$30$^d$ &0.187        &29.76$\pm$0.29    & 657           & F14, G16, R14, L93  \\
    2003hl    &NGC\;0772     &SA(s)b         &52868.5$\pm$5    &1.55*        &32.39$\pm$0.30    &2475           & A14, F14            \\
    2003hn    &NGC\;1448     &SAcd? edge-on  &52866.5$\pm$10   &0.408*       &31.14$\pm$0.40    &1170           & A14, H08, O10       \\
    2004A     &NGC\;6207     &SA(s)c         &53010$\pm$10     &0.180*       &31.44$\pm$0.40    & 852           & T08a, H06, G08, S09, H98       \\
    2004dj    &NGC\;2403     &SAB(s)cd       &53170$\pm$8      &0.107        &27.54$\pm$0.24    & 133           & Z06, T08b, V06, C05, F01, S10  \\
    2004et    &NGC\;6946     &SAB(rs)cd      &53271$\pm$1      &1.3*         &28.67$\pm$0.40    &  40           & S06, F14, B14, E08  \\
    2005ay    &NGC\;3938     &SA(s)c         &53456$\pm$10$^e$ &0.34*        &31.27$\pm$0.30    & 809           & GY08, F14, P09      \\
    2009ib    &NGC\;1559     &SB(s)cd        &55041.3$\pm$3.1  &0.5*         &31.48$\pm$0.30    &1304           & T15, K04            \\
    2012aw    &M\,95         &SB(r)b         &56003            &0.23         &30.00$\pm$0.22    & 778           & D14, M13, B13, V91  \\
    2012ec    &NGC\;1084     &SA(s)c         &56143$\pm$2      &0.31*        &31.33$\pm$0.43    &1407           & S15, B15, Ma13, So14, K04 \\
    2013ej    &M\,74         &SA(s)c         &56500            &0.19         &29.93$\pm$0.12    & 657           & Y16, V14, L93       \\
    2014cx    &NGC\;337      &SB(s)d         &56901$\pm$1.5    &0.297        &31.33$\pm$0.43    &1698           & V16, C16, So14, V91 \\ \hline
  \end{tabular}

  \begin{minipage}{1.0\textwidth}

    $^a$: In most cases, the visual extinction A$_V$ corresponds to the Galactic
    extinction \citep{Schlafly2011}. When the quoted A$_V$ has a
    superscript *, its value corresponds to the total extinction, taken from
    the cited literature.

    $^b:$
    C71: \cite{Ciatti1971},     
    A76: \cite{Arnett1976},     
    L12: \cite{Lennarz2012},    
    T88: \cite{Tifft1988},      
    H01: \cite{Hamuy-phd},      
    A14: \cite{Anderson2014a},  
    O10: \cite{Olivares2010},   
    F96: \cite{Fixsen1996},     
    F14: \cite{Faran2014},      
    P09: \cite{Poznanski2009},  
    V91: \cite{Vaucouleurs1991},
    J09: \cite{Jones2009},      
    S92: \cite{Strauss1992},    
    R14: \cite{Rodriguez2014},  
    S09: \cite{Springob2009},   
    T07: \cite{Theureau2007},   
    G16: \cite{Galbany2016},
    L93: \cite{Lu1993},         
    H08: \cite{Harutyunyan2008},
    T08a: \cite{Tsvetkov2008a},
    H06: \cite{Hendry2006},
    G08: \cite{Gurugubelli2008},
    H98: \cite{Haynes1998},
    Z06: \cite{Zhang2006},
    T08b: \cite{Tsvetkov2008b},
    V06: \cite{Vinko2006},
    C05: \cite{Chugai2005},
    F01: \cite{Freedman2001},
    S10: \cite{Sellwood2010},
    S06: \cite{Sahu2006},
    B14: \cite{Bose2014},
    E08: \cite{Epinat2008},
    GY08: \cite{Gal-Yam2008},
    T15: \cite{Takats2015},
    So14: \cite{Sorce2014},
    K04: \cite{Koribalski2004},
    D14: \cite{DallOra2014},
    M13: \cite{Munari2013},
    B13: \cite{Bose2013},
    B15: \cite{Barbarino2015},
    S15: \cite{Smartt2015},
    Ma13: \cite{Maund2013},
    Y16: \cite{Yuan2016},
    V14: \cite{Valenti2014},
    V16: \cite{Valenti2016},
    C16: \cite{Childress2016}.

    $^c$: For 2003T, we adopt explosion epoch of MJD\;52645 in contrast to MJD\;52654.5
    (A14), basing on the plateau length (around 100 days in our case) and colors
    of spectra.

    $^d$: For 2003gd, we adopt explosion epoch of MJD\;52720$\pm$30, basing on the
    plateau length. The SN was discovered on MJD\;52802, so the explosion epoch
    is highly uncertain.

    $^e$: For 2005ay, we adopt explosion epoch of MJD\;52645, basing on the plateau
    length and colors of spectra.

  \end{minipage}

\end{table*}

\subsection{SN\,1994N}
SN\,1994N was discovered in UGC\,5695 on 10 May 1994 during an observation of the
Type IIn SN\,1993N with the ESO 3.6m telescope \citep{Turatto1994}. For SN\,1994N
we use the photometric and spectroscopic data from \citet{Pastorello2004}.

\subsection{SN\,1997D}

SN\,1997D was discovered in NGC\;1536 on 14 January
1997 about 100 days after maximum \citep{deMello1997}, so the explosion epoch
is not accurately constrained (MJD~$50361\pm15$; \citealt{Spiro2014}).
The red spectra and narrow lines together with the LC indicate that this
object was captured at the end of the plateau phase. In some works a short
plateau of 40--50\,d is proposed \citep{Chugai2000}. Given the high
homogeneity of all the low-luminosity SNe II-P known to date, we adopt in
this work a more conventional plateau length of ${\sim}120$~days, following
\citet{Zampieri2003, Pastorello2004} and \citet{Spiro2014}.
For SN\,1997D, we use the photometric and spectroscopic data from \cite{Benetti2001}.

\subsection{SN\,1999br}
SN\,1999br was discovered in NGC\;4900 on 12 April 1999 \citep{King1999}.
There is no evidence of the SN on frames taken on 4.4 April 1999 with a limiting
magnitude 17 \citep{Yoshida1999}.

While \citep{Pastorello2004} adopts a distance of 17.3~Mpc, \citet{Pignata2013}
argues that this value is probably underestimated based on the
similarities between SN\,1999br and SN\,2008bk. In this paper, we use the
distance of 24.0\,Mpc --- the mean result from the NED catalogue, derived from
seven distance estimations to the host galaxy NGC\;4900.
With this distance, SN\,1999br remains amongst the faintest in our sample of
low-luminosity SNe II-P.

The SN is located on the periphery of NGC\;4900. The spectra do not show any
evidence for significant internal extinction \citep{Pastorello2004}.

For SN\,1999br, we use the photometric and spectroscopic data from \cite{Pastorello2004}.

\subsection{SN\,1999eu}
SN\,1999eu was discovered in NGC\;1097 on 5 November 1999 \citep{Nakano1999a}.
It is located in an arm of the host galaxy.
For SN\,1999eu we use the photometric and spectroscopic data from \cite{Pastorello2004}.

\subsection{SNe~1999gn, 2006ov and 2008in}

The galaxy M\,61 (NGC\;4303) hosts 3 SNe from our sample.

SN\,1999gn was discovered on 17 December 1999 \citep{Dimai1999},
approximately 10 days after explosion.
For SN\,1999gn we use the spectroscopic data from \cite{Spiro2014}.
There are only two $V$-band measurements reported in
\cite{Dimai1999} and \cite{Kiss2000}.

SN\,2006ov was discovered on 24 November 2006 \citep{Nakano2006}.
For SN\,2006ov, we use the photometric and spectroscopic data from \citet{Spiro2014}.

SN\,2008in was discovered on 26 December 2008 \citep{Nakano2008}.
We use the photometric and spectroscopic data from \citet{Roy2011}.
For SN\,2008in, we adopt A$_V$ (estimated as a sum of Galactic and host
galaxy extinction) of 0.305 \citep{Roy2011}.

\subsection{SN\,2001dc}

This SN was discovered  on 30 May 2001 close to the nucleus of the edge-on
Type Sbc galaxy NGC\;5777 \citep{Hurst2001}.
For SN\,2001dc we use the photometric and spectroscopic data from \cite{Spiro2014}.

The position of SN in the host galaxy and its color indicate significant
reddening. We adopt the total extinction $A_V=$\,1.25 \citep{Spiro2014}.

We adopt the distance modulus $\mu = 33.19\pm0.43$  \citep{Sorce2014}
rather than the value $\mu = 32.85$ from LEDA used in \cite{Pastorello2004, Spiro2014}.

\subsection{SN\,2002gd}
SN\,2002gd was discovered in NGC\;7537 on 5 October 2002 \citep{Klotz2002},
probably early after explosion \citep{Spiro2014}.
For SN\,2002gd, we use the photometric and spectroscopic data from \cite{Spiro2014}.

\subsection{SN\,2003Z}
SN\,2003Z was discovered in NGC\;2742 on 29 January 2003 by Qiu \& Hu \citep{Boles2003}.
For SN\,2003Z we use the photometric and spectroscopic data from \cite{Spiro2014}.

\subsection{SN\,2004eg}

SN\,2004eg was discovered in UGC\,3053 on 1 September 2004
\citep{Young2004}. Only two spectra are available at 93 and 171\,d
after the inferred time of explosion.
For SN\,2004eg, we use the photometric and spectroscopic data from
\cite{Spiro2014}. The total extinction $A_V=$\,1.237\,mag \citep{Spiro2014}.

\subsection{SN\,2005cs}

SN\,2005cs was discovered in NGC\;2742 on 30 June 2005 \citep{Modjaz2005}.
For SN\,2005cs, we use the photometric and spectroscopic data from \cite{Pastorello2009}.
We adopt the total extinction $A_V=$\,0.124\,mag as inferred from the multi-epoch
photometric and spectroscopic modeling of \cite{Dessart2008}.

\subsection{SN\,2008bk}

SN\,2008bk was discovered in the NGC\;7793 on 26 March 2008 \citep{Monard2008}.
A low-mass RSG was identified in archival images \citep{Mattila2008, VanDyk2012}.
Modeling of the SN radiation by L17 suggests that SN\,2008bk likely arises
from the low energy explosion of a low-mass RSG star.

For SN\,2008bk, we use the photometric and spectroscopic data from \cite{Pignata2013}
and spectropolarimetric observations from \cite{Leonard2012}.
Following L17, we adopt an extinction $A_V=$\,0.062\,mag, which is within 0.01\,mag
of the value reported by \cite{Schlafly2011} for the line-of-sight towards
the host galaxy NGC\;7793.

\subsection{SN\,2009N}

SN\,2009N was discovered in NGC\;4487 on 24 January 2009 \citep{Nakano2009a}.
For SN\,2009N, we use the photometric and spectroscopic data from \cite{Takats2014}.
The total extinction $A_V=$\,0.403\,mag is estimated from the equivalent width
of the Na\one\,D line by \cite{Takats2014}.

\subsection{SN\,2009md}

SN\,2009md was discovered in NGC\;3389 on 4.81 December 2009 \citep{Nakano2009b}.
For SN\,2009md, we use the photometric and spectroscopic data from \cite{Fraser2011},
where the authors adopted an explosion epoch of $55162\pm8$\,MJD. With
such a choice, the color evolution at early times strongly disagrees with
other objects from our sample. We thus adopted an explosion epoch of
$55170$\,MJD, which is the upper limit of the value adopted in
\cite{Fraser2011}.

\subsection{SN\,2010id}\label{2010id}

SN\,2010id was discovered in NGC\;7483 on 16.33 September 2010 \citep{Lin2010}.
For SN\,2010id, we use the photometric and spectroscopic data from \cite{Gal-Yam2011}.

The $V$-band photometry we use for this SN comes from the Katzman Automatic
Imaging Telescope (KAIT). Its LCs are presented in Fig.~4 of
\cite{Gal-Yam2011}. The $V$-band photometry does not seem to be in good
agreement with other bands: while observations in the $g$ and $r$ bands with
various instruments from 55480 to 55490\,MJD do not show any change in the
slope of the LC, there is an almost 0.5\,mag rise in $V$ as recorded with KAIT.
The photometry has probably some significant errors so we treat all results for
SN\,2010id in Section~\ref{analysis-obs-data} with caution.

\subsection{SN\,2013am}

SN\,2013am was discovered in M65 on 21 March 2013 \citep{Yaron2013}.
For SN\,2013am, we use the photometric and spectroscopic data from \cite{Zhang2014}.
The adopted sum of the Galactic and host galaxy extinction $A_V=$\,1.767 \citep{Zhang2014}.

\subsection{Other candidates in low-luminosity SNe II-P}

In addition to the SNe discussed in the preceding section,  a few more events have been
claimed as low-luminosity Type II-P SNe. However, after analysis of their observational properties,
we decided to exclude them from our sample. These correspond to SNe 1991G, 2003ie and 2014bi.
Below, we explain why they are not considered in the present study.

\subsubsection{SN\,1991G}

Adopting a distance of 15.5\,Mpc and an extinction $A_V=$\,0.025\,mag
(which is lower than the extinction due to Milky Way alone according to
\citealt{Schlegel1998} and \citealt{Schlafly2011}),
for SN\,1991G \citep{Blanton1995},  one would obtain a plateau brightness similar
to that of SN\,2002gd (which is the brightest in our low-luminosity sample).
This seems incompatible with the expansion rate of 6450\,\kms\ inferred
from Fe\two\,5169\,\AA\ at $\sim$\,21 days since explosion,
which is a typical value for standard SNe Type II-P.

With such a low value of the extinction, the first available spectrum
observed on 14 February 1991 ($\sim$\,21\,d after explosion) would be
extremely red in comparison to the other SNe in our sample. On the contrary, if
one adopts a high extinction $A_V$ of 2.0\,mag, the spectrum
has a similar color to those of SNe II-P and SN\,1991G then becomes brighter
than the standard Type II-P SN\,1999em.

Due to these uncertainties with the extinction and/or with the calibration of
the spectra, we do not include SN\,1991G in the current study.

\subsubsection{SN\,2003ie}

SN\,2003ie was classified by \cite{Arcavi2013} as a faint Type II-P event.
However, its intrinsic brightness at discovery is higher than for any
SN in our sample. Moreover, the poor cadence of photometric
observations and the availability of only one spectrum do not allow to
estimate the explosion date. The spectrum is quite different from the spectra
from our sample (e.g., the strength of H$\alpha$ line is lower).
We thus exclude this object from our study.

\subsubsection{SN\,2014bi}

SN\,2014bi has been classified as a faint SN II-P \citep{Zhang2014}.
The observational data are very scarce, only one low-quality spectrum and few
photometric points are available \citep{Zheng2014, Zhang2014}.

Since the inferred $R$-band absolute magnitude is fainter than $-12$\,mag, the
extinction should be very high \citep{Khazov2016}. The exact value of the
extinction is unknown, therefore we cannot determine the SN luminosity.
We thus exclude this object from our study.

\subsection{Type II-P SN\,1999em}
SN\,1999em was discovered in NGC\;1637 on 30 October 1999 by \citet{Li1999}.
SN\,1999em is a standard Type II-P SN with $M_V{\sim}\,-17$\,mag during the
plateau phase.
It has a dense set of spectroscopic and photometric observational data.
We use SN\,1999em as a reference for a standard  SN II-P to compare
to our low-luminosity sample (this does not imply that all standard-luminosity
Type II SNe are like SN\,1999em; see \citealt{Anderson2014a} for a discussion
on the observed diversity of Type II SNe).

The photometric and spectroscopic data are taken from \cite{Hamuy2001} and
\cite{Leonard2003}.
We adopt a total extinction $A_V=$\,0.310\,mag \citep{DH2006}.

\label{lastpage}
\end{document}